%% file: main.tex
\documentclass[paper]{JFM-FLM_Au}
\usepackage{amsmath}
\usepackage{amssymb}
\usepackage{xcolor}
\hypersetup{colorlinks=false,pdfborder={0 0 0}}
\usepackage[T1]{fontenc}
\usepackage[utf8]{inputenc}

\lefttitle{Havlica \textit{et al.}}
\righttitle{Curvature effects on Graetz--L\'ev\^eque transport}

\title{Curvature effects on Graetz--L\'ev\^eque transport from finite wall bands in pipe flow}

\author{Jaromir Havlica\aff{1,2}, Yaroslav Bazaikin\aff{3}, Vaclav Harrandt\aff{1,2}, Florian Huchet\aff{4} \and Jaroslav Tihon\aff{1}}

\affiliation{\aff{1}The Czech Academy of Sciences, Institute of Chemical Process Fundamentals, Rozvojova 135, 165 00 Prague, Czech Republic
\aff{2} Department of Chemistry, Faculty of Science, Jan Evangelista Purkyn{\v e} University in {\'U}st{\'i} nad Labem, Pasteurova 3632/15, 400 96 {\'U}st{\'i} nad Labem, Czech Republic
\aff{3} Department of Mathematics, Faculty of Science, Jan Evangelista Purkyn{\v e} University in {\'U}st{\'i} nad Labem, Pasteurova 3632/15, 400 96 {\'U}st{\'i} nad Labem, Czech Republic
\aff{4}Univ. Gustave Eiffel, MAST-GPEM, Fr-44344 Bouguenais, France}

\makeatletter
\def\@corresau{\resizebox{\linewidth}{!}{\textbf{Corresponding authors:} Yaroslav Bazaikin, yaroslav.bazaykin@ujep.cz; %
  Jaromir Havlica, havlica@icpf.cas.cz}}
\makeatother

\newif\iflegacy
\legacytrue
\input{arxiv-layout}
\begin{document}
\maketitle

\begin{abstract}
We examine heat or mass transfer in fully developed Poiseuille flow through a circular pipe to a finite annular wall band held at a prescribed scalar value and followed by an insulated wall. At fixed wall shear, finite radius reduces transfer over the active band but increases the residual scalar deficit at the insulated wall downstream. Finite pipe radius enters the near-wall problem through two first-order mechanisms: the cylindrical correction to wall-normal diffusion and the quadratic finite-radius correction to the near-wall velocity profile. In the insulated wake, the curvature correction increases the wall scalar deficit: the planar deficit decays with exponent \(-2/3\), whereas the curvature-induced correction decays more slowly, with exponent \(-1/3\). Direct numerical solutions of the finite-radius parabolic boundary-layer problem confirm the corrected wall-flux law and the wake response. For a reference L\'ev\^eque thickness at the band end equal to one-tenth of the pipe radius, the correction reduces the wall-deficit prediction error from \(11.5\,\%\) to \(1.7\,\%\) one band length downstream of the active band. The predictions apply in the high-P\'eclet regime while the scalar boundary layer remains thin compared with the pipe radius.
\end{abstract}

\section{Introduction}
\label{sec:intro}

\par\begingroup\color{black}
We consider heat or mass transfer from fully developed Poiseuille flow in a circular pipe to a finite annular wall band held at a prescribed scalar value. The wall is insulated upstream and downstream of the band, so the fluid-mechanical questions are how finite pipe radius changes the flux on the active band and the residual scalar deficit along the insulated wall.
\par\endgroup

Convective transfer of heat or mass in laminar duct flow is a classical problem in which a wall-transfer law can be derived from the local fluid mechanics rather than introduced empirically. In the Graetz problem, a passive scalar is transported by a fully developed Poiseuille flow while diffusing across the cross-section, with a change in the wall condition imposed at a prescribed axial position \citep{Graetz1885}. At large P\'eclet number, the scalar variation near the entrance of an active region is confined to a thin wall layer. The leading balance is then between advection by the local wall shear and wall-normal diffusion, while axial diffusion is asymptotically smaller. This is the L\'ev\^eque limit \citep{Leveque1928}, which gives the familiar inverse-one-third-power dependence on distance from the leading edge for the local wall flux and, equivalently, for the local Sherwood--Graetz relation. This entrance-layer law remains a basic element of laminar heat- and mass-transfer theory because it captures the singular near-inlet behaviour that complements the downstream Graetz eigenfunction description \citep{ShahLondon1978,Deen2012}.

\par\begingroup\color{black}Wall-transfer measurements give this problem a direct role in fluid mechanics. Wall-flush thermal sensors \citep{BellhouseSchultz1966} and electrodiffusion probes \citep{MitchellHanratty1966} infer wall shear from a measured heat flux or current. For fixed fluid properties, scalar difference and sensor geometry, the steady planar L\'ev\^eque flux is proportional to the cube root of the wall shear rate. A finite-radius change in flux at unchanged wall shear can therefore bias the inferred shear if interpreted using a planar calibration. For successive active regions, the upstream scalar deficit also changes the scalar supply to the next region. The present annular-band solution quantifies the finite-radius flux correction and the downstream wall response needed to model this interaction. An instrument-specific prediction additionally requires the probe geometry, calibration and wall condition on each active region.\par\endgroup The present paper revisits the Graetz--L\'ev\^eque limit from an operator viewpoint. Instead of considering only the classical case in which the wall value is prescribed over a semi-infinite active surface, we regard the wall flux as the image of the wall scalar trace under a Dirichlet--to--Neumann map. In the planar L\'ev\^eque problem this map is nonlocal in the streamwise coordinate. After Laplace transformation in the streamwise coordinate, the wall-normal problem reduces to an Airy equation, and the wall-flux symbol has the characteristic one-third-power dependence on the Laplace variable, a structure already implicit in the analyses of \citet{Lighthill1950} and \citet{Liepmann1958}. In physical space this symbol corresponds to a Riemann--Liouville fractional derivative of order \(1/3\) \citep{OldhamSpanier1974,Podlubny1999}. This formulation is especially useful when the wall condition is mixed: on an active wall region the wall value is prescribed, whereas on an insulated downstream wall the flux is prescribed and the wall value must be determined as part of the solution.

Wall curvature is the additional effect considered here. In a circular pipe, finite radius modifies the near-wall scalar problem in two distinct ways. First, the cylindrical diffusion operator contains a metric correction. Second, the Poiseuille velocity is not exactly linear across the scalar layer: its first finite-radius correction is quadratic in the distance from the wall. Thus the first curvature correction combines a contribution from the cylindrical diffusion metric with a contribution from the quadratic finite-radius correction to the velocity profile. Higher-order L\'ev\^eque expansions for circular tubes and annular passages were developed by \citet{WorsoeSch1967}, and related first correction terms were obtained by \citet{Newman1969}. Heat transfer in nonlinear shear profiles, including quadratic shear, was analysed by \citet{SpenceBrown1968}. These works show that finite-radius corrections for a uniformly active wall are not, by themselves, new. What remains missing is a curvature-corrected wall-flux operator and its use in mixed Dirichlet--Neumann wall problems.

Our main result is such an operator. In the planar L\'ev\^eque limit, the Dirichlet--to--Neumann map has a fractional wall-flux symbol and is therefore nonlocal in the streamwise direction. Weak pipe curvature changes this map in a particularly simple way: at first order it adds a local negative correction to the wall-flux law. The coefficient of this correction is \(-3/5\) in the shear-based scaling used below. This coefficient contains both the cylindrical metric correction to diffusion and the quadratic finite-radius correction to the near-wall Poiseuille velocity. The detailed notation and derivation are given in \S\ref{sec:DtN_Laplace}; the important point here is that the local ordering is controlled by the instantaneous boundary-layer thickness relative to the pipe radius, rather than by a global curvature parameter alone.

For a uniformly active wall, this correction leaves the classical L\'ev\^eque entrance exponent unchanged and adds a constant negative correction to the wall flux. In the local, diameter-based Sherwood expansion used here, this appears as the constant term \(-6/5\). This result recovers the classical extended L\'ev\^eque correction for a circular tube in the present Sherwood--Graetz convention \citep{WorsoeSch1967,Newman1969}. It is therefore a consistency check, not the main novelty of the present work. The more consequential effect appears when the wall condition is mixed. For a finite annular active band followed by an insulated wall, the zero-flux condition in the wake must be imposed through the nonlocal Dirichlet--to--Neumann map. The planar wake trace decays downstream with exponent \(-2/3\), whereas the first curvature correction decays with exponent \(-1/3\). Thus a local correction to the wall-flux operator is converted, by the mixed-boundary closure, into a slowly decaying downstream wake response. This finite-radius change in scalar recovery is the main physical result identified in this paper.

Mixed active--passive wall conditions occur whenever transfer is localised to finite wall regions, for example in heated or reactive strips and in heterogeneous reactive--inert surfaces \citep{AckerbergPatelGupta1978,ShahLinShaqfeh2017}. They also arise in electrodiffusion wall-shear measurements, where the measured current is a wall flux and segmented probes introduce active electrodes separated by insulating gaps \citep{SonHanratty1969,WeinWichterle1989,Sobolik1990,HavlicaKramolis2021}. Recent probe models address finite electrode geometry, arbitrary flow direction and multi-segment interactions \citep{HarrandtHavlica2023,HarrandtHavlica2024,HarrandtHavlica2025}. These examples motivate the mixed-boundary setting considered below, although the present pipe model is axisymmetric and corresponds to an annular active band rather than to an azimuthally localised electrode. In a companion planar analysis \citep{HarrandtHavlica2026}, Laplace-transform methods were used to couple the active segment to the insulated wake in the planar problem. The present work extends that framework to weakly curved pipe walls and shows how curvature modifies both the local transfer coefficient and the downstream wake. \par\begingroup\color{black}The laminar calculation provides a controlled reference for interpreting wall-transfer measurements. In turbulent flow at high Schmidt number, mass transfer can probe velocity variations within the viscous sublayer \citep{SirkarHanratty1970}. For fully developed turbulent flow in a smooth pipe, the mean velocity has the same quadratic near-wall term as the Poiseuille profile at the same mean wall shear. This follows from the mean momentum balance because the Reynolds shear stress vanishes cubically with distance from a smooth impermeable wall. The connection extends beyond the locally linear profile, but does not by itself establish the accuracy of the present correction for turbulent scalar transport. Such an application requires the neglected turbulent transport to be assessed relative to the retained finite-radius correction, and fluctuating measurements additionally require the temporal response of the scalar layer \citep{Huchet2011}.\par\endgroup

The paper is organised as follows. Section~\ref{sec:phys_config} formulates the finite-radius Graetz--L\'ev\^eque boundary-layer problem and identifies the small parameters controlling the reduction. Section~\ref{sec:DtN_Laplace} derives the curvature-corrected Dirichlet--to--Neumann map in Laplace and physical space. Section~\ref{sec:uniform_wall} applies this map to the uniformly active wall and obtains the constant curvature correction to the local flux and the corresponding Sherwood-number shift. Section~\ref{sec:single_electrode} applies the same map to a finite annular active band followed by an insulated wake. Section~\ref{sec:results} compares the asymptotic predictions with direct numerical solutions of the finite-radius boundary-layer problem and discusses the range of validity of the weak-curvature theory.

\section{Finite-radius Graetz--L\'ev\^eque formulation}
\label{sec:phys_config}

We consider steady transport of a passive scalar in fully developed laminar flow through a circular pipe of radius \(R\). Cylindrical coordinates \((r,\theta,z)\) are used, with \(z\) measured downstream and the wall located at \(r=R\). The flow and scalar field are taken axisymmetric. Consequently, the active region considered below is an annular wall band, uniform in the azimuthal direction; azimuthal edge effects of localised wall probes are outside the present model. The velocity field is the Poiseuille profile
\begin{equation}
  u_z(r)=U_{\mathrm{c}}\left(1-\frac{r^2}{R^2}\right),
  \qquad
  U_{\mathrm{b}}:=\frac{2}{R^2}\int_0^R u_z(r)r\,\mathrm{d}r
  =
  \frac{U_{\mathrm{c}}}{2},
  \label{eq:uz_profile_dim}
\end{equation}
where \(U_{\mathrm{c}}\) is the centreline velocity and \(U_{\mathrm{b}}\) is the cross-sectional mean velocity. The scalar diffusivity is denoted by \(D\), and the scalar field \(\phi(r,z)\) satisfies
\begin{equation}
  u_z(r)\frac{\partial\phi}{\partial z}
  =
  D\left[
    \frac{1}{r}\frac{\partial}{\partial r}
    \left(r\frac{\partial\phi}{\partial r}\right)
    +
    \frac{\partial^2\phi}{\partial z^2}
  \right].
  \label{eq:adv_diff_cyl}
\end{equation}
The inlet scalar value is uniform,
\begin{equation}
  \phi(r,0)=\phi_\infty,
  \qquad 0\le r\le R,
  \label{eq:inlet_bc}
\end{equation}
and the active wall value, when imposed, is denoted by \(\phi_{\mathrm{w}}\).

\subsection{Pipe problem and near-wall reduction}
\label{sec:nearwall_scaling}

The analysis concerns the high-P\'eclet Graetz--L\'ev\^eque limit near the pipe wall. We introduce the wall-normal coordinate
\begin{equation}
  x=R-r,
  \qquad x\ge 0,
\end{equation}
so that \(x=0\) is the wall and \(x=R\) is the pipe axis. In this coordinate the cylindrical diffusion operator is exactly
\begin{equation}
  \frac{1}{r}\frac{\partial}{\partial r}\left(r\frac{\partial\phi}{\partial r}\right)
  =
  \frac{\partial^2\phi}{\partial x^2}
  -
  \frac{1}{R-x}\frac{\partial\phi}{\partial x}.
  \label{eq:diff_dim_exact_x_nw}
\end{equation}
The wall shear rate is
\begin{equation}
  S_{\mathrm{w}}
  :=
  \left.\frac{\mathrm{d}u_z}{\mathrm{d}x}\right|_{x=0}
  =
  -\left.\frac{\mathrm{d}u_z}{\mathrm{d}r}\right|_{r=R}
  =
  \frac{2U_{\mathrm{c}}}{R}
  =
  \frac{4U_{\mathrm{b}}}{R},
  \label{eq:Sw_def}
\end{equation}
and the Poiseuille velocity has the exact near-wall form
\begin{equation}
  u_z(R-x)
  =
  S_{\mathrm{w}} x
  -
  \frac{S_{\mathrm{w}}}{2R}x^2
  =
  S_{\mathrm{w}} x\left(1-\frac{x}{2R}\right).
  \label{eq:uz_nearwall_dim_nw}
\end{equation}
The leading term \(S_{\mathrm{w}}x\) is the planar linear-shear contribution. The term proportional to \(x^2/R\) is the first finite-radius correction to the near-wall velocity profile. Thus finite radius enters the boundary-layer problem in two places: through the cylindrical metric term in \eqref{eq:diff_dim_exact_x_nw} and through this quadratic velocity correction.

In the near-wall layer, \(x\sim\delta(z)\) and \(z\sim \ell_*\), where \(\ell_*\) is a reference streamwise length. Balancing shear advection with wall-normal diffusion gives the L\'ev\^eque thickness
\begin{equation}
  \delta_{\mathrm{b}}
  =
  \left(\frac{D\ell_*}{S_{\mathrm{w}}}\right)^{1/3}.
  \label{eq:delta_b_def}
\end{equation}
Axial diffusion is smaller than wall-normal diffusion by \(O(\beta^2)\), where
\begin{equation}
  \beta:=\frac{\delta_{\mathrm{b}}}{\ell_*}.
  \label{eq:beta_def}
\end{equation}
The present analysis takes the parabolic Graetz--L\'ev\^eque limit \(\beta\ll1\) and neglects axial diffusion away from the axial-diffusion edge layers generated by abrupt changes in wall condition. The reduced Graetz boundary-layer equation is therefore
\begin{equation}
  u_z(R-x)\frac{\partial\phi}{\partial z}
  =
  D\left[
    \frac{\partial^2\phi}{\partial x^2}
    -
    \frac{1}{R-x}\frac{\partial\phi}{\partial x}
  \right].
  \label{eq:graetz_dimensional_nw}
\end{equation}
This parabolic approximation is used throughout the analysis below. The full axial-diffusion problem \eqref{eq:adv_diff_cyl} is used only to identify the \(\beta\ll1\) reduction and is not solved as part of the main theory.

\subsection{Shear scaling, wall data and flux}
\label{sec:shear_nondim}

We non-dimensionalise the near-wall problem using \(\ell_*\) and \(\delta_{\mathrm{b}}\):
\begin{equation}
  Z=\frac{z}{\ell_*},
  \qquad
  X=\frac{x}{\delta_{\mathrm{b}}},
  \qquad
  \Theta(X,Z)
  =
  \frac{\phi(r,z)-\phi_{\mathrm{w}}}{\phi_\infty-\phi_{\mathrm{w}}},
  \qquad
  \varepsilon=\frac{\delta_{\mathrm{b}}}{R}.
  \label{eq:nd_vars_BL}
\end{equation}
Applying this scaling to the parabolic finite-radius equation \eqref{eq:graetz_dimensional_nw} gives
\begin{equation}
  X\left(1-\frac{\varepsilon X}{2}\right)
  \frac{\partial\Theta}{\partial Z}
  =
  \frac{\partial^2\Theta}{\partial X^2}
  -
  \frac{\varepsilon}{1-\varepsilon X}
  \frac{\partial\Theta}{\partial X},
  \qquad
  0<X<\varepsilon^{-1},
  \quad Z>0.
  \label{eq:graetz_BL_XZ}
\end{equation}
For comparison, if the axial-diffusion term in the full scalar equation \eqref{eq:adv_diff_cyl} is retained before taking the parabolic limit, the same scaling gives
\begin{equation}
  X\left(1-\frac{\varepsilon X}{2}\right)
  \frac{\partial\Theta}{\partial Z}
  =
  \frac{\partial^2\Theta}{\partial X^2}
  -
  \frac{\varepsilon}{1-\varepsilon X}
  \frac{\partial\Theta}{\partial X}
  +
  \beta^2
  \frac{\partial^2\Theta}{\partial Z^2}.
  \label{eq:full_AD_scaled}
\end{equation}
Equation \eqref{eq:graetz_BL_XZ} is obtained from \eqref{eq:full_AD_scaled} by setting \(\beta=0\). Thus \(\varepsilon=\delta_{\mathrm{b}}/R\) and \(\beta=\delta_{\mathrm{b}}/\ell_*\) measure two different approximations: \(\varepsilon\) measures weak curvature within the parabolic boundary-layer model, whereas \(\beta\) measures the neglect of axial diffusion in passing from the full scalar advection--diffusion equation to the parabolic Graetz--L\'ev\^eque model. The analysis below sets \(\beta=0\) and expands in \(\varepsilon\).

The pipe axis is located at \(X=\varepsilon^{-1}\). In the asymptotic analysis, \(X=O(1)\) and \(Z=O(1)\) are held fixed while \(\varepsilon\to0\), so the axis recedes to infinity and the domain is replaced by the half-line \(X\in(0,\infty)\). For the finite-radius numerical benchmark in \S\ref{sec:results}, the interval \(0<X<\varepsilon^{-1}\) is retained and the symmetry condition is imposed at the axis.

The inlet and outer conditions for the boundary-layer problem are
\begin{equation}
  \Theta(X,0)=1,
  \qquad X>0,
  \label{eq:bc_inlet_BL}
\end{equation}
and
\begin{equation}
  \Theta(X,Z)\to1
  \qquad \text{as} \qquad X\to\infty
  \quad \text{for each fixed } Z>0.
  \label{eq:bc_outer_BL}
\end{equation}
The dimensionless wall flux is
\begin{equation}
  J(Z)
  :=
  \left.\frac{\partial\Theta}{\partial X}\right|_{X=0},
  \label{eq:wall_flux_def_BL}
\end{equation}
and the corresponding dimensional flux is
\begin{equation}
  j(z)
  =
  -D\left.\frac{\partial\phi}{\partial r}\right|_{r=R}
  =
  \frac{D(\phi_\infty-\phi_{\mathrm{w}})}{\delta_{\mathrm{b}}}\,J(Z).
  \label{eq:wall_flux_scaling_BL}
\end{equation}

\begin{figure}
\centering
\includegraphics[width=\textwidth]{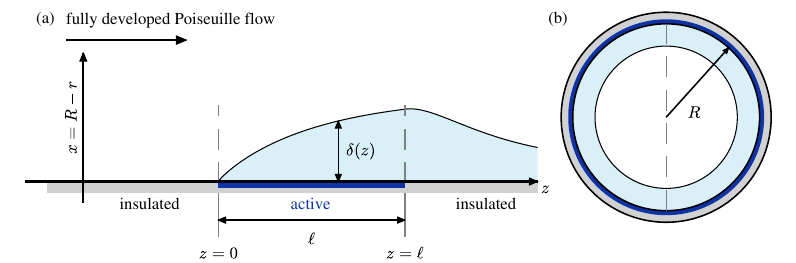}
\caption{Schematic of the finite-radius Graetz--L\'ev\^eque problem. (\emph{a}) Local near-wall representation of the pipe wall. The wall is insulated upstream, active over the annular band \(0<z<\ell\), and insulated again downstream. The solid blue strip marks the axial extent of the active band in the near-wall representation. The light-blue filled region above the wall represents the scalar boundary layer and the downstream residual wake: it starts at the upstream edge of the active band, \(z=0\), grows along the active wall and relaxes over the insulated wall for \(z>\ell\). The wall-normal coordinate is \(x=R-r\). The drawn thickness of the light-blue region is schematic and not drawn to scale. (\emph{b}) Pipe cross-section through the active band. The blue inner circumference denotes that the active wall condition is uniform in the azimuthal direction; the finite axial extent of the active band is shown in panel (\emph{a}).}
\label{fig:problem_schematic}
\end{figure}

It is useful to introduce the wall trace
\begin{equation}
  G(Z)
  :=
  1-\Theta(0,Z).
  \label{eq:G_walltrace_BL}
\end{equation}
On an active wall, \(\Theta(0,Z)=0\), equivalently \(G(Z)=1\). For the uniformly active wall,
\begin{equation}
  \Theta(0,Z)=0,
  \qquad Z>0.
  \label{eq:bc_uniform_wall_BL}
\end{equation}
For a finite annular active band, the upstream edge may be placed at \(z=0\) without loss of generality in the parabolic Graetz model: an upstream insulated wall leaves the uniform inlet state unchanged. If the active length is \(\ell\), we choose \(\ell_*=\ell\), so that the active band occupies \(0<Z<1\). The mixed wall condition is then written compactly as
\begin{subequations}
\label{eq:bc_wall_mixed_BL}
\begin{alignat}{2}
  G(Z)&=1, &\qquad& 0<Z<1, \\
  J(Z)&=0, &\qquad& Z>1.
\end{alignat}
\end{subequations}
The finite-band configuration is sketched in figure~\ref{fig:problem_schematic}: the scalar boundary layer starts at the upstream edge of the active band, grows over \(0<z<\ell\), and leaves a residual wake over the downstream insulated wall.

The immediate neighbourhoods of the leading edge \(Z=0\) and the switching point \(Z=1\) are interpreted within the same parabolic boundary-layer model. In the full scalar problem with \(\beta>0\), axial diffusion would regularise abrupt changes in wall condition in thin edge layers; this edge-layer structure is outside the present approximation. {A numerical comparison retaining axial diffusion is reported in \S\ref{sec:results_decay}.}

Finally, the diameter-based Sherwood number is
\begin{equation}
  \mathrm{Sh}_D(Z)
  =
  \frac{2R\,j(z)}{D(\phi_\infty-\phi_{\mathrm{w}})}
  =
  \frac{2}{\varepsilon}J(Z).
  \label{eq:ShD_def}
\end{equation}
The conventional diameter-based Graetz number is
\begin{equation}
  \mathrm{Gz}_D(z)
  =
  \frac{U_{\mathrm{b}}(2R)^2}{Dz}
  =
  \frac{S_{\mathrm{w}}R^3}{Dz}
  =
  \frac{\varepsilon^{-3}}{Z}.
  \label{eq:GzD_def}
\end{equation}
The local L\'ev\^eque thickness is \(\delta(z)=(Dz/S_{\mathrm{w}})^{1/3}\), and therefore
\begin{equation}
  \frac{\delta(z)}{R}
  =
  \mathrm{Gz}_D(z)^{-1/3}
  =
  \varepsilon Z^{1/3}.
  \label{eq:delta_over_R_Gz}
\end{equation}
We denote this local curvature parameter by
\begin{equation}
  \chi(Z)
  :=
  \frac{\delta(z)}{R}
  =
  \varepsilon Z^{1/3}.
  \label{eq:chi_def}
\end{equation}
The asymptotic theory below therefore assumes both \(\beta=\delta_{\mathrm{b}}/\ell_*\ll1\), which justifies the parabolic reduction away from axial-diffusion edge layers, and \(\chi(Z)\ll1\), which orders the curvature expansion.

\section{Curvature-corrected Dirichlet--to--Neumann map}
\label{sec:DtN_Laplace}

The finite-radius boundary-layer equation \eqref{eq:graetz_BL_XZ} is linear, so the relation between the wall scalar trace and the wall flux can be written as a Dirichlet--to--Neumann (DtN) map. This section derives that map directly. The uniformly active wall is then obtained in \S\ref{sec:uniform_wall} as a simple corollary.

We introduce the deviation from the upstream value,
\begin{equation}
  \vartheta(X,Z):=1-\Theta(X,Z),
  \label{eq:vartheta_def_DtN}
\end{equation}
so that \(\vartheta(X,0)=0\), \(\vartheta(X,Z)\to0\) as \(X\to\infty\), and the wall trace is
\begin{equation}
  G(Z)=\vartheta(0,Z)=1-\Theta(0,Z).
\end{equation}
The dimensionless wall flux is
\begin{equation}
  J(Z)
  =
  \left.\partial_X\Theta\right|_{X=0}
  =
  -\left.\partial_X\vartheta\right|_{X=0}.
  \label{eq:J_def_DtN}
\end{equation}
Thus the aim is to determine the linear operator \(\mathcal{D}_\varepsilon\) such that
\begin{equation}
  J=\mathcal{D}_\varepsilon G.
  \label{eq:DtN_operator_def}
\end{equation}

Taking the Laplace transform in \(Z\),
\begin{equation}
  \widehat{\vartheta}(X,p)
  =
  \int_0^\infty e^{-pZ}\vartheta(X,Z)\,\mathrm{d}Z,
  \qquad
  \widehat G(p)
  =
  \int_0^\infty e^{-pZ}G(Z)\,\mathrm{d}Z,
  \qquad p>0,
  \label{eq:Laplace_def_DtN}
\end{equation}
and using \(\vartheta(X,0)=0\), the transformed boundary-layer equation becomes
\begin{equation}
  \widehat{\vartheta}_{XX}
  -
  \frac{\varepsilon}{1-\varepsilon X}\widehat{\vartheta}_X
  -
  pX\left(1-\frac{\varepsilon X}{2}\right)\widehat{\vartheta}
  =
  0.
  \label{eq:DtN_ODE_exact}
\end{equation}
The asymptotic DtN map is constructed on the near-wall half-line \(0<X<\infty\), where the transformed disturbance satisfies \(\widehat{\vartheta}(X,p)\to0\) as \(X\to\infty\). The pipe axis lies at \(X=\varepsilon^{-1}\). When the Laplace-domain layer thickness \(p^{-1/3}\) satisfies \(\varepsilon p^{-1/3}\ll1\), the decaying wall-normal solution is confined to \(X\ll\varepsilon^{-1}\), so the axis condition does not enter the algebraic weak-curvature expansion.

By linearity we write
\begin{equation}
  \widehat{\vartheta}(X,p)=\widehat G(p)\,\Psi(X,p;\varepsilon),
  \qquad
  \Psi(0,p;\varepsilon)=1,
  \qquad
  \Psi(X,p;\varepsilon)\to0
  \quad \text{as } X\to\infty.
  \label{eq:U_normalised}
\end{equation}
The Laplace transform of the wall flux is then
\begin{equation}
  \widehat J(p;\varepsilon)
  =
  -\left.\partial_X\widehat{\vartheta}\right|_{X=0}
  =
  \Lambda(p;\varepsilon)\widehat G(p),
  \label{eq:DtN_symbolic}
\end{equation}
where
\begin{equation}
  \Lambda(p;\varepsilon)
  =
  -\left.\partial_X\Psi(X,p;\varepsilon)\right|_{X=0}
  \label{eq:Lambda_def}
\end{equation}
is the Laplace-domain DtN symbol.

\subsection{Planar L\'ev\^eque map}
\label{sec:DtN_planar}

For \(\varepsilon=0\), equation \eqref{eq:DtN_ODE_exact} reduces to
\begin{equation}
  \Psi_{0,XX}-pX\Psi_0=0,
  \qquad
  \Psi_0(0,p)=1,
  \qquad
  \Psi_0(X,p)\to0
  \quad \text{as } X\to\infty.
  \label{eq:U0_ODE}
\end{equation}
With \(\xi=p^{1/3}X\), \eqref{eq:U0_ODE} becomes the Airy equation
\begin{equation}
  \frac{\mathrm{d}^2\Psi_0}{\mathrm{d}\xi^2}
  -
  \xi\Psi_0
  =
  0.
  \label{eq:Psi0_Airy_equation}
\end{equation}
This Airy reduction is the standard local form of the planar L\'ev\^eque problem \citep{Lighthill1950,Liepmann1958}. The solution \(\mathrm{Bi}(\xi)\) grows as \(\xi\to\infty\), whereas \(\mathrm{Ai}(\xi)\) decays. Normalisation by \(\Psi_0(0,p)=1\) therefore gives
\begin{equation}
  \Psi_0(X,p)
  =
  \frac{\mathrm{Ai}\!\left(p^{1/3}X\right)}{\mathrm{Ai}(0)}.
  \label{eq:U0_Ai}
\end{equation}
Consequently,
\begin{equation}
  \Lambda_0(p)
  =
  -\Psi_{0,X}(0,p)
  =
  \kappa_0p^{1/3},
  \qquad
  \kappa_0
  :=
  -\frac{\mathrm{Ai}'(0)}{\mathrm{Ai}(0)}
  =
  3^{1/3}\frac{\Gamma(2/3)}{\Gamma(1/3)}.
  \label{eq:Lambda0_symbol}
\end{equation}
The planar DtN relation is thus
\begin{equation}
  \widehat J_0(p)=\kappa_0p^{1/3}\widehat G(p).
  \label{eq:DtN_planar_hat}
\end{equation}
For causal wall traces in the present entrance problem, multiplication by \(p^{1/3}\) in Laplace space corresponds to the Riemann--Liouville derivative of order \(1/3\), so that
\begin{equation}
  J_0(Z)
  =
  \kappa_0\mathcal{D}_Z^{1/3}G(Z),
  \label{eq:DtN_planar_phys}
\end{equation}
with
\begin{equation}
  \mathcal{D}_Z^{1/3}G(Z)
  =
  \frac{1}{\Gamma(2/3)}
  \frac{\mathrm{d}}{\mathrm{d}Z}
  \int_0^Z (Z-S)^{-1/3}G(S)\,\mathrm{d}S.
  \label{eq:RL_13_def_Z}
\end{equation}
The planar wall-flux operator is therefore nonlocal in the streamwise coordinate.

\subsection{First curvature correction}
\label{sec:DtN_first_order}

We now expand the symbol for weak curvature. For \(X=O(1)\), expansion of \eqref{eq:DtN_ODE_exact} gives
\begin{equation}
  \Psi_{XX}-pX\Psi
  +
  \varepsilon\left(
    \frac{pX^2}{2}\Psi-\Psi_X
  \right)
  =
  O(\varepsilon^2).
  \label{eq:DtN_ODE_expanded}
\end{equation}
Writing
\begin{equation}
  \Psi(X,p;\varepsilon)
  =
  \Psi_0(X,p)
  +
  \varepsilon \Psi_1(X,p)
  +
  O(\varepsilon^2),
  \label{eq:U_expansion}
\end{equation}
we obtain
\begin{equation}
  \Psi_{1,XX}-pX\Psi_1
  =
  \Psi_{0,X}
  -
  \frac{pX^2}{2}\Psi_0,
  \qquad
  \Psi_1(0,p)=0,
  \qquad
  \Psi_1(X,p)\to0
  \quad \text{as } X\to\infty.
  \label{eq:U1_ODE}
\end{equation}
Introduce \(\xi=p^{1/3}X\) and write \(\Psi_1=p^{-1/3}W_1(\xi)\). Then \eqref{eq:U1_ODE} becomes
\begin{equation}
  W_1''-\xi W_1
  =
  \frac{1}{\mathrm{Ai}(0)}
  \left[
    \mathrm{Ai}'(\xi)-\frac{\xi^2}{2}\mathrm{Ai}(\xi)
  \right],
  \qquad
  W_1(0)=0,
  \qquad
  W_1(\xi)\to0
  \quad \text{as } \xi\to\infty.
  \label{eq:W1_ODE}
\end{equation}
A decaying solution satisfying \(W_1(0)=0\) can be sought as a polynomial combination of \(\mathrm{Ai}\) and \(\mathrm{Ai}'\). Using \(\mathrm{Ai}''(\xi)=\xi\mathrm{Ai}(\xi)\), direct substitution into \eqref{eq:W1_ODE} verifies that
\begin{equation}
  W_1(\xi)
  =
  \frac{1}{\mathrm{Ai}(0)}
  \left[
    \frac{3}{5}\xi\mathrm{Ai}(\xi)
    -
    \frac{1}{10}\xi^2\mathrm{Ai}'(\xi)
  \right].
  \label{eq:W1_explicit}
\end{equation}
Since \(\Psi_{1,X}(0,p)=W_1'(0)=3/5\), the first curvature correction to the DtN symbol is independent of \(p\). Hence
\begin{equation}
  \Lambda(p;\varepsilon)
  =
  \kappa_0p^{1/3}
  -
  \frac{3}{5}\varepsilon
  +
  O(\varepsilon^2p^{-1/3}),
  \qquad
  \varepsilon p^{-1/3}\ll1.
  \label{eq:Lambda_asymp}
\end{equation}
Equivalently,
\begin{equation}
  \frac{\Lambda(p;\varepsilon)}{p^{1/3}}
  =
  \kappa_0
  -
  \frac{3}{5}\varepsilon p^{-1/3}
  +
  O\!\left((\varepsilon p^{-1/3})^2\right).
  \label{eq:Lambda_scaled_asymp}
\end{equation}
This form makes explicit that the small parameter in the Laplace-domain boundary layer is the ratio of the layer thickness \(p^{-1/3}\) to the radius \(1/\varepsilon\).

The coefficient \(-3/5\) has two physically distinct contributions. Solving the two forced Airy problems separately, the forcing term \(\Psi_{0,X}\) in \eqref{eq:U1_ODE}, which comes from the cylindrical diffusion operator, gives \(W_{1,\mathrm{diff}}'(0)=1/2\). The forcing term \(-pX^2\Psi_0/2\), which comes from the quadratic correction to the Poiseuille velocity, gives \(W_{1,\mathrm{vel}}'(0)=1/10\). Since the flux symbol is \(-\Psi_X(0)\), these become
\begin{equation}
  -\frac{3}{5}
  =
  -\frac{1}{2}
  -
  \frac{1}{10},
  \label{eq:minus_three_fifths_split}
\end{equation}
\par\begingroup\color{black}
The two contributions have the same sign: each reduces the flux associated with a prescribed positive wall deficit relative to the planar linear-shear result. Their magnitudes are different. The cylindrical diffusion metric contributes \(-1/2\) and the quadratic velocity term contributes \(-1/10\), giving the total \(-3/5\) in \eqref{eq:minus_three_fifths_split}. Their shares of this coefficient are therefore \((-1/2)/(-3/5)=5/6\) and \((-1/10)/(-3/5)=1/6\), respectively. The velocity contribution compares Poiseuille flow with a linear profile having the same wall shear rate: at a finite distance from the wall, the quadratic term reduces the axial speed and changes the balance between advection and diffusion across the scalar layer. Thus retaining the exact near-wall velocity while using planar diffusion would recover only one-sixth of the first-order correction. Both terms are needed to describe transport along the inner wall of a circular pipe.
\par\endgroup

\par\begingroup\color{black}
The quadratic velocity term is fixed by the pressure-driven momentum balance. For the present fully developed Newtonian flow, write \(u(x)=u_z(R-x)\), with \(x=R-r\), and denote the pressure and dynamic viscosity by \(P\) and \(\mu\). Here primes denote differentiation with respect to the inward wall-normal distance \(x\); \(u''(0)\) is the second derivative of the axial velocity profile at the wall. The axial momentum balance at the wall gives
\[
  \mu\left[u''(0)-\frac{S_{\mathrm{w}}}{R}\right]
  =\frac{\mathrm{d}P}{\mathrm{d}z}.
\]
For Poiseuille flow, \(\mathrm{d}P/\mathrm{d}z=-2\mu S_{\mathrm{w}}/R\), hence \(u''(0)=-S_{\mathrm{w}}/R\), as in \eqref{eq:uz_nearwall_dim_nw}. The velocity contribution \(-1/10\) therefore reflects the profile curvature determined jointly by the pressure gradient and pipe geometry, whereas the \(-1/2\) contribution arises from cylindrical scalar diffusion. This connection between pressure gradient, quadratic velocity profile and wall transfer is related to the analysis of \citet{SpenceBrown1968}.
\par\endgroup

\subsection{Physical-space and dimensional forms}
\label{sec:DtN_dimensional}

Inverting \eqref{eq:Lambda_asymp} gives the curvature-corrected DtN map
\begin{equation}
  J(Z;\varepsilon)
  =
  \kappa_0\mathcal{D}_Z^{1/3}G(Z)
  -
  \frac{3}{5}\varepsilon G(Z)
  +
  \mathcal{R}_2[G](Z),
  \label{eq:DtN_phys_eps}
\end{equation}
where the remainder is to be understood at the symbol level as
\begin{equation}
  \widehat{\mathcal{R}}_2(p)
  =
  O(\varepsilon^2p^{-1/3})\widehat G(p),
  \qquad
  \varepsilon p^{-1/3}\ll1.
  \label{eq:DtN_remainder_symbol}
\end{equation}
The key structural point is that the planar operator is fractional and nonlocal, whereas the first curvature correction is local in physical space.

Returning to dimensional variables, let
\begin{equation}
  \widetilde G(z):=G\!\left(\frac{z}{\ell_*}\right).
\end{equation}
The scaling property of the Riemann--Liouville derivative gives
\begin{equation}
  \left(\mathcal{D}_Z^{1/3}G\right)\!\left(\frac{z}{\ell_*}\right)
  =
  \ell_*^{1/3}\mathcal{D}_z^{1/3}\widetilde G(z).
  \label{eq:RL_scaling_property}
\end{equation}
Using \(\delta_{\mathrm{b}}=(D\ell_*/S_{\mathrm{w}})^{1/3}\), the dimensional wall flux becomes
\begin{equation}
  j(z)
  =
  (\phi_\infty-\phi_{\mathrm{w}})
  \left[
    \kappa_0(D^2S_{\mathrm{w}})^{1/3}\mathcal{D}_z^{1/3}\widetilde G(z)
    -
    \frac{3}{5}\frac{D}{R}\widetilde G(z)
  \right]
  +
  \frac{D(\phi_\infty-\phi_{\mathrm{w}})}{\delta_{\mathrm{b}}}
  \mathcal{R}_2[G]\!\left(\frac{z}{\ell_*}\right).
  \label{eq:dim_DtN_eps}
\end{equation}
For wall traces varying on the reference scale \(\ell_*\), the retained curvature term is \(O(D/R)\), while the neglected term is smaller by a factor \(O(\delta_{\mathrm{b}}/R)\).

\section{Uniformly active wall and Sherwood-number shift}
\label{sec:uniform_wall}

The semi-infinite uniformly active wall, starting at the leading edge \(Z=0\), is obtained by setting
\begin{equation}
  G(Z)=1,
  \qquad Z>0.
  \label{eq:uniform_G}
\end{equation}
The planar part of the DtN map gives
\begin{equation}
  \mathcal{D}_Z^{1/3}1
  =
  \frac{1}{\Gamma(2/3)}Z^{-1/3},
\end{equation}
and therefore
\begin{equation}
  J_0(Z)
  =
  \frac{\kappa_0}{\Gamma(2/3)}Z^{-1/3}
  =
  \frac{3^{1/3}}{\Gamma(1/3)}Z^{-1/3}.
  \label{eq:J0_planar_uniform}
\end{equation}
This is the classical leading-order local L\'ev\^eque flux law.

Including the first curvature correction from \eqref{eq:DtN_phys_eps} yields
\begin{equation}
  J(Z;\varepsilon)
  =
  \frac{3^{1/3}}{\Gamma(1/3)}Z^{-1/3}
  -
  \frac{3}{5}\varepsilon
  +
  O(\varepsilon^2Z^{1/3}),
  \qquad
  \varepsilon Z^{1/3}\ll1.
  \label{eq:J_uniform_final}
\end{equation}
Thus weak curvature does not change the L\'ev\^eque entrance exponent. It adds a constant negative correction to the dimensionless wall flux. The relative correction scales as \(\varepsilon Z^{1/3}=\delta(z)/R\), as expected from the local thickness ratio.

The corresponding dimensional transfer coefficient \(k(z)=j(z)/(\phi_\infty-\phi_{\mathrm{w}})\) is
\begin{equation}
  k(z)
  =
  \frac{3^{1/3}}{\Gamma(1/3)}
  \left(\frac{D^2S_{\mathrm{w}}}{z}\right)^{1/3}
  -
  \frac{3}{5}\frac{D}{R}
  +
  O\!\left(
    \frac{D}{R}\frac{\delta(z)}{R}
  \right),
  \qquad
  \delta(z)=\left(\frac{Dz}{S_{\mathrm{w}}}\right)^{1/3}.
  \label{eq:kdim_uniform}
\end{equation}
For the local, diameter-based Sherwood number defined with the inlet/outer scalar difference,
\begin{equation}
  \mathrm{Sh}_D(Z)
  =
  \frac{2R\,j(z)}{D(\phi_\infty-\phi_{\mathrm{w}})}
  =
  \frac{2}{\varepsilon}J(Z),
\end{equation}
we obtain
\begin{equation}
  \mathrm{Sh}_D
  =
  \frac{2\,3^{1/3}}{\Gamma(1/3)}
  \frac{R}{\delta(z)}
  -
  \frac{6}{5}
  +
  O\!\left(\frac{\delta(z)}{R}\right).
  \label{eq:ShD_uniform}
\end{equation}
Equivalently, using \(\delta(z)/R=\mathrm{Gz}_D^{-1/3}\),
\begin{equation}
  \mathrm{Sh}_D
  =
  \frac{2\,3^{1/3}}{\Gamma(1/3)}
  \mathrm{Gz}_D^{1/3}
  -
  \frac{6}{5}
  +
  O\!\left(\mathrm{Gz}_D^{-1/3}\right).
  \label{eq:ShD_uniform_Gz}
\end{equation}
The constant term \(-6/5\) is the uniform-wall manifestation of the local curvature correction in the present Sherwood--Graetz convention. It provides a consistency check against the extended L\'ev\^eque expansion for a circular tube \citep{WorsoeSch1967,Newman1969} and gives a transparent interpretation of the operator correction. The main advantage of the DtN form appears in the mixed wall problem considered next, where the same local term is converted into a downstream wake response.

\section{Finite annular active band and insulated wake}
\label{sec:single_electrode}

We now apply the curvature-corrected Dirichlet--to--Neumann map to a finite annular active wall band followed by an insulated downstream wall. This mixed boundary condition is the setting in which the operator formulation is most useful: the wall value is prescribed on the active part, but must be determined in the wake from the zero-flux condition.

The upstream edge of the annular active band is placed at \(z=0\), and the reference axial length is chosen as the active length, \(\ell_*=\ell\). Thus the active band occupies \(0<Z<1\), and the wake corresponds to \(Z>1\). In terms of the wall trace \(G(Z)=1-\Theta(0,Z)\) and wall flux \(J(Z)=\Theta_X(0,Z)\), the wall conditions are
\begin{subequations}
\label{eq:finite_segment_wall_conditions}
\begin{align}
  G(Z)&=1,
  &&0<Z<1, \\
  J(Z)&=0,
  &&Z>1.
\end{align}
\end{subequations}
The unknown in the wake is therefore \(G(Z)\) for \(Z>1\).

\subsection{Flux on the active band}
\label{sec:single_electrode_flux}

The DtN map \eqref{eq:DtN_phys_eps} is causal in \(Z\). Hence, for \(0<Z<1\), the downstream insulated wall does not affect the local flux in the parabolic Graetz--L\'ev\^eque model. Since \(G=1\) on the active band, \eqref{eq:J_uniform_final} gives
\begin{equation}
  J(Z;\varepsilon)
  =
  \frac{3^{1/3}}{\Gamma(1/3)}Z^{-1/3}
  -
  \frac{3}{5}\varepsilon
  +
  O(\varepsilon^2Z^{1/3}),
  \qquad
  0<Z<1,
  \qquad
  \varepsilon Z^{1/3}\ll1.
  \label{eq:single_electrode_flux_dimless}
\end{equation}
In dimensional form,
\begin{equation}
  j(z)
  =
  (\phi_\infty-\phi_{\mathrm{w}})
  \left[
    \frac{3^{1/3}}{\Gamma(1/3)}
    (D^2S_{\mathrm{w}})^{1/3}z^{-1/3}
    -
    \frac{3}{5}\frac{D}{R}
  \right]
  +
  (\phi_\infty-\phi_{\mathrm{w}})
  O\!\left(
    \frac{D}{R}\frac{\delta(z)}{R}
  \right),
  \qquad
  0<z<\ell.
  \label{eq:single_electrode_flux_dimensional}
\end{equation}
Thus the finite active band has the same local curvature correction as the uniformly active wall, away from the switching point at \(Z=1\). The difference between the two problems appears downstream, where the wall trace is no longer prescribed.

\subsection{Planar wake}
\label{sec:single_electrode_planar_wake}

In the wake, the zero-flux condition must be imposed through the DtN map. We expand the wall trace as
\begin{equation}
  G(Z)
  =
  G_0(Z)+\varepsilon G_1(Z)+O(\varepsilon^2),
  \qquad
  Z>1,
  \label{eq:g_expansion_section5}
\end{equation}
with \(G_0=1\) on \(0<Z<1\) and \(G_1=0\) on \(0<Z<1\), because the active-wall Dirichlet condition is imposed exactly.

At leading order, the zero-flux condition gives
\begin{equation}
  \mathcal{D}_Z^{1/3}G_0(Z)=0,
  \qquad
  Z>1,
  \label{eq:wake_planar_fractional}
\end{equation}
where the Riemann--Liouville derivative includes the history \(G_0=1\) on \(0<Z<1\). Equivalently,
\begin{equation}
  \int_1^Z (Z-S)^{-1/3}G_0(S)\,\mathrm{d}S
  =
  \frac{3}{2}
  \left[
    1+(Z-1)^{2/3}-Z^{2/3}
  \right],
  \qquad
  Z>1.
  \label{eq:Abel_single_electrode_planar}
\end{equation}
Solving this Abel equation gives the planar wake trace obtained by \citet{HarrandtHavlica2026}
\begin{equation}
  G_0(Z)
  =
  \frac{
    B_{1/Z}\!\left(\frac{2}{3},\frac{1}{3}\right)
  }{
    B\!\left(\frac{2}{3},\frac{1}{3}\right)
  },
  \qquad
  Z\ge1,
  \label{eq:g0_beta}
\end{equation}
where \(B_x(a,b)\) is the lower incomplete beta function. This solution satisfies \(G_0(1)=1\) and \(G_0(Z)\to0\) as \(Z\to\infty\). Its far-wake behaviour is
\begin{equation}
  G_0(Z)
  \sim
  \frac{3\sqrt{3}}{4\pi}Z^{-2/3},
  \qquad
  Z\to\infty.
  \label{eq:g0_asymptotic}
\end{equation}

\subsection{Curvature correction to the wake}
\label{sec:single_electrode_wake_eps}

At first order in curvature, the zero-flux condition \(J=0\) and the DtN map \eqref{eq:DtN_phys_eps} give
\begin{equation}
  \kappa_0\mathcal{D}_Z^{1/3}G_1(Z)
  =
  \frac{3}{5}G_0(Z),
  \qquad
  Z>1,
  \label{eq:g1_fractional_forcing}
\end{equation}
with \(G_1=0\) on \(0<Z<1\). Thus the local curvature correction in the wall-flux operator becomes a nonlocal fractional-integration problem in the wake.

Introducing \(t=Z-1\) and \(H_i(t)=G_i(1+t)\), equation \eqref{eq:g1_fractional_forcing} gives, after Laplace transformation in \(t\),
\begin{equation}
  \widehat H_1(p)
  =
  \frac{3}{5\kappa_0}p^{-1/3}\widehat H_0(p).
  \label{eq:g1_Laplace_relation}
\end{equation}
Therefore
\begin{equation}
  G_1(Z)
  =
  \frac{3}{5\kappa_0\Gamma(1/3)}
  \int_1^Z
    (Z-S)^{-2/3}G_0(S)\,\mathrm{d}S,
  \qquad
  Z>1.
  \label{eq:g1_fractional_integral}
\end{equation}
This expression is the most transparent form of the curvature correction: \(G_1\) is a fractional integral of the planar wake.

The far-wake behaviour follows from \eqref{eq:g0_beta} and \eqref{eq:g1_fractional_integral}:
\begin{equation}
  G_0(Z)\sim \frac{3\sqrt{3}}{4\pi}Z^{-2/3},
  \qquad
  G_1(Z)\sim
  \frac{9}{10\kappa_0\Gamma(2/3)^2}Z^{-1/3},
  \qquad
  Z\to\infty.
  \label{eq:wake_asymptotics}
\end{equation}
The first curvature correction therefore decays more slowly than the planar wake. Consequently,
\begin{equation}
  \frac{\varepsilon G_1(Z)}{G_0(Z)}
  =
  O\!\left(\varepsilon Z^{1/3}\right)
  =
  O\!\left(\frac{\delta(z)}{R}\right).
  \label{eq:wake_relative_curvature}
\end{equation}
The same local parameter \(\delta(z)/R\) that controls the uniform-wall expansion also controls the wake expansion, but in the wake it appears as a growing relative correction because \(G_1\) decays more slowly than \(G_0\). The slower decay should therefore not be extrapolated to distances for which \(\varepsilon Z^{1/3}=O(1)\); it indicates the increasing importance of curvature up to the edge of the weak-curvature regime, not a breakdown of the local ordering within its range of validity.

Defining the wall scalar value as \(\Theta_{\mathrm{w}}(Z)=\Theta(0,Z)\), the wake value is obtained from \(\Theta_{\mathrm{w}}=1-G\):
\begin{equation}
  \Theta_{\mathrm{w}}(Z)
  =
  1-G_0(Z)-\varepsilon G_1(Z)
  +
  O(\varepsilon^2),
  \qquad
  Z>1.
  \label{eq:Theta_w_wake_expansion}
\end{equation}

\par\begingroup\color{black}
\subsection{Physical interpretation of the wake scalings}
\label{sec:wake_physical_interpretation}

The wake exponents can be understood from conservation of scalar
deficit and the spreading of the scalar layer. With
\(\vartheta=1-\Theta\), integration of the parabolic equation
\eqref{eq:graetz_BL_XZ} gives
\begin{equation}
\begin{aligned}
\mathcal M_\varepsilon(Z)
&=
\int_0^{1/\varepsilon}
X\left(1-\frac{\varepsilon X}{2}\right)
(1-\varepsilon X)\,\vartheta(X,Z)\,\mathrm dX,
\\
\frac{\mathrm d\mathcal M_\varepsilon}{\mathrm dZ}
&=J(Z).
\end{aligned}
\label{eq:wake_deficit_balance}
\end{equation}
Here \(\mathcal M_\varepsilon\) is the dimensionless advective flux
of scalar deficit; its weights account for the velocity and the
annular cross-sectional area. In the insulated wake, \(J=0\), so
this flux remains constant at the value supplied by uptake on
the active band.

In the planar limit, the balance between shear advection and
transverse diffusion gives a dimensionless layer thickness
\(L\propto Z^{1/3}\). On the active band, this yields the familiar
flux scaling \(J_0\propto L^{-1}\). In the far wake, write
\(\vartheta_0\sim G_0(Z)F(\eta)\), where \(\eta=X/L\),
\(F(0)=1\), and \(F\) is the limiting scaled profile.
Conservation then gives
\begin{equation}
\mathcal M_0
\sim
G_0(Z)L^2
\int_0^\infty \eta F(\eta)\,\mathrm d\eta
=\mathrm{constant},
\qquad
G_0\propto L^{-2}\propto Z^{-2/3}.
\label{eq:wake_planar_physical_scaling}
\end{equation}
One factor of \(L\) comes from the layer width and the other
from the linear velocity profile. Recovery at the wall therefore
reflects redistribution of a conserved deficit flux across
a growing layer.

Finite radius changes how this conserved flux is carried across the
layer. At the same wall shear and the same distance from the wall,
the Poiseuille profile gives a lower axial velocity than a planar
linear-shear profile. The area available for axial transport is also
smaller than in the planar approximation because the circumference
decreases towards the pipe axis.
The combined weight in \eqref{eq:wake_deficit_balance} is
\(X(1-3\varepsilon X/2)+O(\varepsilon^2X^3)\).
As the layer spreads, this weight reduces the flux carried by an
unchanged deficit profile by a relative amount of order
\(\varepsilon L\). Conservation therefore requires a redistribution
of the deficit of that same relative order.

This redistribution and the change in total uptake have different
downstream scalings. With \(L=Z^{1/3}\), the planar similarity profile
is \(F(\eta)=\exp(-\eta^3/9)\). Writing
\(\vartheta=\vartheta_0+\varepsilon\vartheta_1+\cdots\), the first
two far-wake terms of the curvature correction have the form
\[
\varepsilon\vartheta_1
\sim \varepsilon Z^{-1/3}g(\eta)
     +\varepsilon a_1Z^{-2/3}F(\eta).
\]
Here \(g\) describes redistribution across the growing layer, whereas
\(a_1\) accounts for the first-order change in uptake on the active band.
That change is negative, \(\mathcal M_1=-3/5\), from
\eqref{eq:single_electrode_flux_dimless}, and contributes only to the
faster-decaying term. The slower term is thus not driven by the
first-order reduction in uptake, although its amplitude depends on
the leading planar uptake. Appendix~\ref{sec:appendix_wake_similarity}
derives these terms from the scalar equation and the conserved flux,
and recovers the wall amplitude in \eqref{eq:wake_asymptotics}.
The positive kernel in \eqref{eq:g1_fractional_integral} gives
\(G_1>0\). A smaller integrated uptake is therefore consistent with a
larger local wall deficit: the two quantities describe different
aspects of the redistributed scalar field.

The relative curvature effect grows downstream as the layer thickens.
These power laws describe the intermediate range
\(1\ll Z\ll\varepsilon^{-3}\), while the layer remains thin
compared with the pipe radius.
For comparison, complete transverse mixing in the finite-radius
parabolic problem would leave a uniform deficit
\(G_\infty=8\varepsilon^2\mathcal M_\varepsilon\), obtained by integrating
the weight in \eqref{eq:wake_deficit_balance}.
This finite cross-sectional capacity is consistent with increasing
relative importance of curvature, but complete mixing lies outside
the thin-layer expansion and is not used to infer the sign or amplitude
of its first correction.
\par\endgroup

\par\begingroup\color{black}
The practical consequence is that a second active band or wall sensor receives a different scalar supply from that predicted by a planar model. For a scalar-absorbing band held at the same wall value, the larger incoming deficit reduces the local driving difference available for further transfer. The planar model already describes depletion by the upstream band; the present correction quantifies the additional finite-radius change. Interpreting that additional change in heat flux or current using a planar calibration could therefore bias the inferred wall shear, even when the shear is unchanged. The present result supplies the downstream wall response needed to model this interaction; the actual flux on a second band must be calculated with its own active-wall condition imposed. These conclusions apply while the local scalar-layer thickness remains small relative to the pipe radius, as specified in \eqref{eq:wake_relative_curvature}.\par\endgroup

\par\begingroup\color{black}
The size of this effect can be expressed directly in terms of the local
layer-to-radius ratio. Equations \eqref{eq:g0_beta} and
\eqref{eq:g1_fractional_integral} give
\(\varepsilon G_1/G_0\simeq0.73\,\delta(z)/R\),
\(0.88\,\delta(z)/R\) and \(1.14\,\delta(z)/R\) at
\(Z=1.5\), \(2\) and \(5\), respectively. The coefficient approaches
\(1.63\) in the far-wake asymptotics \eqref{eq:wake_asymptotics},
within \(1\ll Z\ll\varepsilon^{-3}\). Here \(z\) is measured
from the leading edge of the active band. A parameter example in
\S\ref{sec:results_validity} translates these ratios into percentage
changes in the wall deficit.
\par\endgroup

\section{Validation and physical consequences}
\label{sec:results}

\par\begingroup\color{black}The preceding sections give an asymptotic wall-flux operator for weakly curved Graetz--L\'ev\^eque boundary layers in the parabolic limit \(\beta=0\). We first compare its physical consequences with direct numerical solutions of the finite-radius cylindrical boundary-layer problem \eqref{eq:graetz_BL_XZ}: the local wall flux, the Sherwood-number shift and the wake trace \(G(Z)=1-\Theta(0,Z)\) behind a finite annular active band. These comparisons test the curvature expansion within the same high-P\'eclet Graetz approximation from which it is derived. The separate Laplace-domain test of the Dirichlet--to--Neumann symbol follows after the physical-space results. The boundary-layer benchmarks do not validate axial-diffusion effects in the full advection--diffusion equation \eqref{eq:adv_diff_cyl}.\par\endgroup

\par\begingroup\color{black}
The comparisons test wall flux and downstream scalar deficit. For a wall held at a constant scalar value for all \(Z>0\), setting \(\varepsilon=0\) recovers the planar L\'ev\^eque result \(J_0=(3^{1/3}/\Gamma(1/3))Z^{-1/3}\): the local flux decreases as the scalar boundary layer grows downstream from the leading edge. Figure~\ref{fig:uniform_validation} compares this flux with the finite-radius prediction and its first correction. Converting the flux correction to the local diameter-based Sherwood number gives the known constant shift \(-6/5\), equal to Newman's \(-1.2\) constant \citep{Newman1967Report}, tested in figure~\ref{fig:sherwood_shift}; appendix~\ref{sec:appendix_extended_leveque} specifies the inlet/outer scalar reference and the conversion. For a finite band followed by an insulated wall, setting \(\varepsilon=0\) instead recovers the planar downstream wall deficit \(G_0\). Figure~\ref{fig:wake_validation} tests the improvement obtained by retaining the first curvature term. The first two comparisons check the classical L\'ev\^eque flux and its known pipe correction; the third tests the mixed active--insulated configuration. The additional Laplace-domain test in figure~\ref{fig:DtN_symbol_validation} checks the coefficient in the wall-flux map independently of either wall configuration.
\par\endgroup

\subsection{Finite-radius boundary-layer benchmark}
\label{sec:results_benchmark}

The parabolic boundary-layer comparisons use direct numerical solutions of \eqref{eq:graetz_BL_XZ} on the finite interval \(0<X<\varepsilon^{-1}\), retaining the finite-radius curvature terms in both advection and diffusion. The wall condition is imposed at \(X=0\), while the pipe-axis symmetry condition is imposed at \(X=\varepsilon^{-1}\). Thus the benchmark does not use the half-line approximation \(X\in(0,\infty)\) used in the asymptotic derivation.

The wall-normal diffusion operator is discretised in conservative form,
\begin{equation}
  \Theta_{XX}
  -
  \frac{\varepsilon}{1-\varepsilon X}\Theta_X
  =
  \frac{1}{1-\varepsilon X}
  \frac{\partial}{\partial X}
  \left[
    (1-\varepsilon X)\Theta_X
  \right],
  \label{eq:benchmark_conservative_operator_results}
\end{equation}
and the solution is advanced implicitly in \(Z\). For the uniformly active wall, \(\Theta(0,Z)=0\) is imposed for all \(Z>0\). For the finite annular active band, the wall condition switches from \(G=1\) on \(0<Z<1\) to \(J=0\) on \(Z>1\).

The computations reported below used \(N_X=1200\) cell-centred wall-normal points. The uniformly active-wall calculations used \(1200\) logarithmically spaced axial points on \(10^{-3}\le Z\le1\). The finite active-band calculations used \(1200\) logarithmically spaced points on \(10^{-3}\le Z\le1\) and \(1800\) linearly spaced points on \(1<Z\le5\). \par\begingroup\color{black}
An additional independent resolution check for the uniformly active wall is reported in Appendix~\ref{sec:appendix_benchmark}. It separates wall-normal, axial and wall-flux-extraction sensitivities. An analogous check for the finite-band wake is reported there separately.
\par\endgroup Further numerical details are given in Appendix~\ref{sec:appendix_benchmark}.

\subsection{Uniformly active wall}
\label{sec:results_uniform_validation}

For a uniformly active wall, the asymptotic prediction is
\begin{equation}
  J(Z;\varepsilon)
  =
  J_0(Z)-\frac{3}{5}\varepsilon
  +
  O(\varepsilon^2Z^{1/3}),
  \qquad
  J_0(Z)=\frac{3^{1/3}}{\Gamma(1/3)}Z^{-1/3}.
  \label{eq:results_uniform_prediction}
\end{equation}
This result tests the local part of the curvature-corrected DtN map. It also tests the predicted ordering of the remainder, since the relative curvature effect is controlled by
\begin{equation}
  \chi(Z)=\varepsilon Z^{1/3}=\frac{\delta(z)}{R}.
  \label{eq:results_chi}
\end{equation}

For the error plots we use the finite-radius boundary-layer solution as the
reference and define
\begin{equation}
  E_J^{(0)}
  =
  \frac{|J_{\mathrm{num}}-J_0|}{J_{\mathrm{num}}},
  \qquad
  E_J^{(1)}
  =
  \frac{|J_{\mathrm{num}}-(J_0-3\varepsilon/5)|}{J_{\mathrm{num}}}.
  \label{eq:uniform_error_def}
\end{equation}
Here \(E_J^{(0)}\) is the error of the planar L\'ev\^eque approximation, whereas \(E_J^{(1)}\) is the error after including the first curvature correction.

\begin{figure}
\centering
\includegraphics[width=\textwidth]{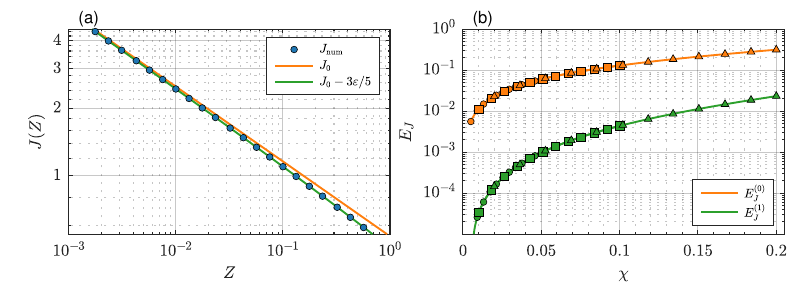}
\caption{Uniformly active wall. All quantities are dimensionless. (\emph{a}) Wall flux \(J(Z)\) for \(\varepsilon=0.10\), comparing the finite-radius cylindrical boundary-layer benchmark \(J_{\mathrm{num}}\) with the planar L\'ev\^eque prediction \(J_0=(3^{1/3}/\Gamma(1/3))Z^{-1/3}\) and the first-order curvature-corrected approximation \(J_0-3\varepsilon/5\). (\emph{b}) Absolute relative errors \(E_J^{(0)}\) and \(E_J^{(1)}\), defined in \eqref{eq:uniform_error_def}, plotted against \(\chi=\varepsilon Z^{1/3}=\delta(z)/R\). Line styles distinguish \(E_J^{(0)}\) and \(E_J^{(1)}\); circles, squares and triangles denote \(\varepsilon=0.05\), \(0.10\) and \(0.20\), respectively.}
\label{fig:uniform_validation}
\end{figure}

Figure~\ref{fig:uniform_validation}(a) compares the finite-radius benchmark with the planar L\'ev\^eque prediction \(J_0\) and the corrected expression \eqref{eq:results_uniform_prediction} for \(\varepsilon=0.10\). The planar model systematically overpredicts the flux. This sign is consistent with the analytical correction: finite radius reduces the wall flux relative to the locally planar approximation. At \(Z\approx0.02\), \(0.10\) and \(1\), the planar relative errors are approximately \(3.1\,\%\), \(5.5\,\%\) and \(13.0\,\%\), respectively. The first-order corrected model reduces these errors to approximately \(0.03\,\%\), \(0.08\,\%\) and \(0.44\,\%\).

Figure~\ref{fig:uniform_validation}(b) plots \(E_J^{(0)}\) and \(E_J^{(1)}\) against the local curvature parameter \(\chi=\varepsilon Z^{1/3}=\delta(z)/R\) for \(\varepsilon=0.05\), \(0.10\) and \(0.20\). For the uniformly active wall, rescaling the wall-normal coordinate by the local L\'ev\^eque thickness shows that \(Z^{1/3}J(Z;\varepsilon)\) depends on \(Z\) and \(\varepsilon\) only through the combination \(\chi\). Consequently, the relative errors are functions of \(\chi\) alone, and the data for different \(\varepsilon\) collapse onto the same two curves over their overlapping \(\chi\)-ranges. The planar error increases with \(\chi\), whereas the corrected approximation remains substantially more accurate over the displayed weak-curvature range. For each value of \(\varepsilon\), the corresponding data terminate at \(\chi=\varepsilon\), because the plotted interval ends at \(Z=1\).

\subsection{Sherwood-number shift}
\label{sec:results_sherwood}

The same result can be expressed in terms of the local, diameter-based Sherwood number defined with the inlet/outer scalar difference. From \eqref{eq:ShD_uniform_Gz},
\begin{equation}
  \mathrm{Sh}_D
  =
  \frac{2\,3^{1/3}}{\Gamma(1/3)}\mathrm{Gz}_D^{1/3}
  -
  \frac{6}{5}
  +
  O(\mathrm{Gz}_D^{-1/3}).
  \label{eq:results_sherwood_prediction}
\end{equation}
Thus the first curvature correction appears as the constant term \(-6/5\) in this Sherwood--Graetz convention.

\begin{figure}
\centering
\includegraphics[width=0.63\textwidth]{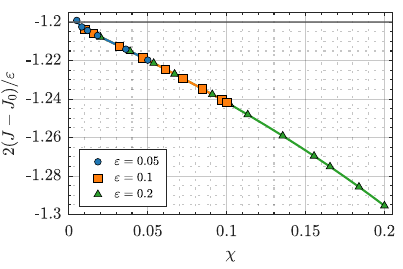}
\caption{Diameter-based Sherwood-number shift for the uniformly active wall. The extracted quantity \(2(J_{\mathrm{num}}-J_0)/\varepsilon\) is plotted against \(\chi=\varepsilon Z^{1/3}=\delta(z)/R\) for \(\varepsilon=0.05\), \(0.10\) and \(0.20\). The dashed horizontal line denotes \(-6/5\).}
\label{fig:sherwood_shift}
\end{figure}

Figure~\ref{fig:sherwood_shift} tests this prediction by plotting \(2(J_{\mathrm{num}}-J_0)/\varepsilon\) against \(\chi\). In the asymptotic range the numerical curves approach the predicted value \(-6/5\). Deviations at larger \(\chi\) reflect the \(O(\varepsilon^2Z^{1/3})\) remainder in the wall flux, or equivalently the \(O(\delta(z)/R)\) remainder in the Sherwood number. The figure therefore confirms both the magnitude and the sign of the local curvature correction in the observable transfer coefficient.

\subsection{Finite annular active band and wake trace}
\label{sec:results_wake}

\begin{figure}[h!]
\centering
\includegraphics[width=\textwidth]{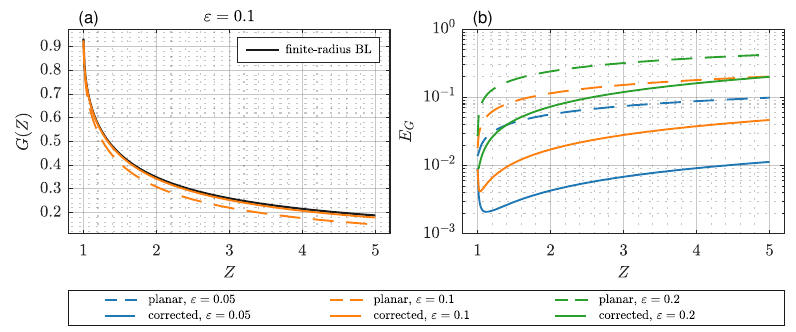}
\caption{Finite annular active band followed by an insulated wake. (\emph{a}) Wake trace \(G(Z)=1-\Theta(0,Z)\) for \(\varepsilon=0.10\), comparing the finite-radius benchmark \(G_{\mathrm{num}}\), the planar wake \(G_0\), and the first-order corrected approximation \(G_0+\varepsilon G_1\). (\emph{b}) Absolute relative errors \(E_G^{(0)}\) and \(E_G^{(1)}\), defined in \eqref{eq:wake_error_def}, for \(\varepsilon=0.05\), \(0.10\) and \(0.20\).}
\label{fig:wake_validation}
\end{figure}

We now consider the mixed boundary condition \(G=1\) on \(0<Z<1\) and \(J=0\) on \(Z>1\), corresponding to a finite annular active band followed by an insulated wall. This case tests the nonlocal mixed-boundary closure. The planar wake trace is \(G_0\), given by \eqref{eq:g0_beta}; the first curvature correction is \(G_1\), given by the fractional integral \eqref{eq:g1_fractional_integral}. The predicted wake trace is therefore
\begin{equation}
  G(Z;\varepsilon)
  =
  G_0(Z)+\varepsilon G_1(Z)+O(\varepsilon^2),
  \qquad
  Z>1.
  \label{eq:results_wake_prediction}
\end{equation}
For the wake-error plots we use
\begin{equation}
  E_G^{(0)}
  =
  \frac{|G_{\mathrm{num}}-G_0|}{G_{\mathrm{num}}},
  \qquad
  E_G^{(1)}
  =
  \frac{|G_{\mathrm{num}}-(G_0+\varepsilon G_1)|}{G_{\mathrm{num}}}.
  \label{eq:wake_error_def}
\end{equation}
Figure~\ref{fig:wake_validation}(a) compares \eqref{eq:results_wake_prediction} with the finite-radius benchmark for \(\varepsilon=0.10\). The planar wake underpredicts \(G(Z)\), while the corrected approximation follows the numerical curve much more closely. At \(Z=1.5\), \(2\) and \(5\), the planar relative errors are approximately \(8.7\,\%\), \(11.5\,\%\) and \(20.2\,\%\), respectively. The first-order corrected approximation reduces these discrepancies to approximately \(1.1\,\%\), \(1.7\,\%\) and \(4.7\,\%\).
Figure~\ref{fig:wake_validation}(b) shows \(E_G^{(0)}\) and \(E_G^{(1)}\) for \(\varepsilon=0.05\), \(0.10\) and \(0.20\). The first-order correction improves the wake prediction for all three values. The remaining error grows more rapidly for \(\varepsilon=0.20\), especially farther downstream, which is consistent with the loss of ordering when \(\varepsilon Z^{1/3}\) is no longer small.

\subsection{Slow downstream decay of the curvature correction}
\label{sec:results_decay}

The stronger downstream sensitivity in the wake follows from the different decay rates of \(G_0\) and \(G_1\):
\begin{equation}
  G_0(Z)\sim \frac{3\sqrt{3}}{4\pi}Z^{-2/3},
  \qquad
  G_1(Z)\sim
  \frac{9}{10\kappa_0\Gamma(2/3)^2}Z^{-1/3},
  \qquad
  Z\to\infty.
  \label{eq:results_wake_asymptotics}
\end{equation}
Thus the first curvature correction decays more slowly than the planar wake. Although it is formally multiplied by \(\varepsilon\), its relative importance grows like \(\varepsilon Z^{1/3}\). This slower decay should not be extrapolated beyond the weak-curvature range; it indicates increasing downstream sensitivity to curvature while \(\varepsilon Z^{1/3}\ll1\).

\begin{figure}
\centering
\includegraphics[width=\textwidth]{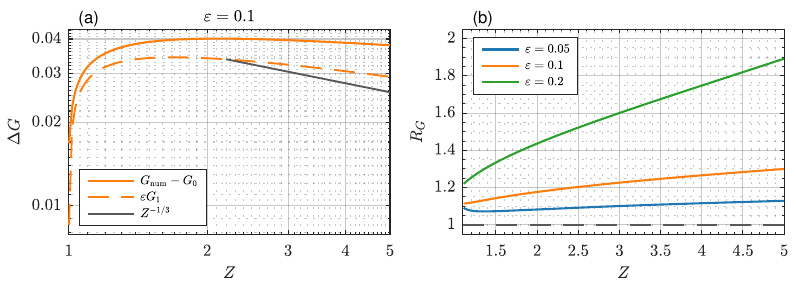}
\caption{Curvature correction to the wake trace. (\emph{a}) Numerically obtained correction \(G_{\mathrm{num}}-G_0\) and first-order prediction \(\varepsilon G_1\) for \(\varepsilon=0.10\). The dotted line indicates the reference decay \(Z^{-1/3}\). (\emph{b}) Ratio diagnostic \(R_G(Z)\), defined in \eqref{eq:results_wake_ratio}, for \(\varepsilon=0.05\), \(0.10\) and \(0.20\).}
\label{fig:wake_decay}
\end{figure}

Figure~\ref{fig:wake_decay}(a) compares the numerically obtained correction \(G_{\mathrm{num}}-G_0\) with the predicted first-order correction \(\varepsilon G_1\) for \(\varepsilon=0.10\). The comparison captures the sign, magnitude and downstream trend of the curvature contribution. Figure~\ref{fig:wake_decay}(b) plots the ratio
\begin{equation}
  R_G(Z)
  :=
  \frac{G_{\mathrm{num}}(Z)-G_0(Z)}{\varepsilon G_1(Z)}.
  \label{eq:results_wake_ratio}
\end{equation}
For weak curvature this ratio remains close to unity over the plotted wake interval. The departure for \(\varepsilon=0.20\) is consistent with the fact that higher-order curvature terms are no longer negligible when the local parameter \(\varepsilon Z^{1/3}\) becomes too large.

\par\begingroup\color{black}
The interval shown in figure~\ref{fig:wake_decay} tests the finite-distance correction given by \eqref{eq:g1_fractional_integral}. The \(Z^{-1/3}\) line is an asymptotic reference; agreement with the correction over this interval does not by itself establish its limiting exponent. That exponent follows from \eqref{eq:wake_asymptotics}.
\par\endgroup

\par\begingroup\color{black}
To assess the effect of retaining axial diffusion, we also solved the full scaled equation \eqref{eq:full_AD_scaled} numerically for one representative case, \(\varepsilon=0.1\) and \(\beta=0.05\), retaining the axial-diffusion term. The wall condition changes at \(Z=1\) as in the parabolic calculation. A uniform inlet is imposed at \(Z=0\), the pipe axis is a symmetry boundary, and the axial derivative is set to zero at \(Z=8\). The full-equation and \(\beta=0\) traces were computed with the same conservative discretisation on identical grids. With \(N_X=200\) and \(\Delta Z=0.00125\), the full-equation wake trace exceeds the \(\beta=0\) trace by approximately \(0.40\,\%\) and \(0.21\,\%\) at \(Z=1.5\) and \(2\), respectively. At \(Z=5\), the magnitude of the difference is below \(0.1\,\%\) on all the tested grids. At these three interior comparison points, the differences remain below \(6\,\%\) of the retained first curvature correction \(\varepsilon G_1\) across the tested grids. Retaining axial diffusion therefore changes the wall deficit by much less than the curvature correction for these boundary conditions. The comparison includes axial diffusion throughout the domain and does not isolate the switching-layer contribution. Moving the outlet from \(Z=8\) to \(12\) at fixed \(N_X=200\) and \(\Delta Z=0.00125\) changed the full-equation values at these points by less than \(10^{-8}\,\%\). Halving this axial step changed the three full-equation values by \(0.0146\,\%\), \(0.0090\,\%\) and \(0.0029\,\%\); doubling the wall-normal resolution at \(\Delta Z=0.00125\) changed them by \(0.0175\,\%\), \(0.0140\,\%\) and \(0.0114\,\%\). The discrete axial scalar-flux balance was conserved to within \(2.0\times10^{-12}\) in the insulated wake. In contrast, the value at \(Z=1.001\) remains sensitive to axial refinement and is not used to quantify the far-wake amplitude. This calculation illustrates the \(\beta=0.05\) case with the stated inlet and outlet conditions; it is not a uniform error bound for arbitrary \(\beta\) or an inlet placed farther upstream.
\par\endgroup
\subsection{Numerical check of the Laplace-domain Dirichlet--to--Neumann symbol}
\label{sec:results_DtN_symbol}

Finally, we check the Laplace-domain symbol of the Dirichlet--to--Neumann map itself. In this context, the symbol is the multiplier \(\Lambda(p;\varepsilon)\) in the relation \(\widehat J(p)=\Lambda(p;\varepsilon)\widehat G(p)\). The check is not a direct numerical simulation in physical space; it is a numerical solution of the finite-radius wall-normal boundary-value problem at fixed Laplace parameter, followed by extraction of \(\Lambda(p;\varepsilon)\) from the wall derivative. It therefore tests \eqref{eq:Lambda_asymp} directly, before any inverse transform or mixed-boundary closure is used.

For this purpose we solve the finite-radius Laplace-domain wall-normal problem without using the weak-curvature perturbation expansion. Introducing the scaled coordinate \(y=p^{1/3}X\) and the Laplace-domain curvature parameter
\begin{equation}
  q=\varepsilon p^{-1/3},
  \label{eq:q_symbol_validation}
\end{equation}
for \(q>0\), the normalised wall-normal problem defining the DtN symbol can be written in conservative form as
\begin{equation}
  \frac{\mathrm d}{\mathrm dy}
  \left[
    (1-qy)\frac{\mathrm d\Psi}{\mathrm dy}
  \right]
  -
  y\left(1-\frac{qy}{2}\right)(1-qy)\Psi
  =0,
  \qquad
  0<y<\frac{1}{q},
  \label{eq:DtN_symbol_validation_ODE}
\end{equation}
with boundary conditions
\begin{equation}
  \Psi(0)=1,
  \qquad
  \Psi_y(1/q)=0.
  \label{eq:DtN_symbol_validation_BC}
\end{equation}
We denote the scaled finite-radius DtN symbol by
\(\Lambda_{\mathrm{s}}(q):=\Lambda(p;\varepsilon)/p^{1/3}\). The numerical approximation obtained from
\eqref{eq:DtN_symbol_validation_ODE}--\eqref{eq:DtN_symbol_validation_BC}
is
\begin{equation}
  \Lambda_{\mathrm{s},\mathrm{num}}(q)
  :=
  \frac{\Lambda_{\mathrm{num}}(p;\varepsilon)}{p^{1/3}}
  =
  -\Psi_y(0).
  \label{eq:scaled_symbol_num}
\end{equation}
The asymptotic prediction \eqref{eq:Lambda_scaled_asymp} becomes
\begin{equation}
  \Lambda_{\mathrm{s}}(q)
  =
  \kappa_0-\frac{3}{5}q+O(q^2),
  \qquad
  q\ll1.
  \label{eq:scaled_symbol_prediction}
\end{equation}
The value at \(q=0\) is obtained from the planar half-line Airy problem and is \(\Lambda_{\mathrm{s}}(0)=\kappa_0\).

\begin{figure}
\centering
\includegraphics[width=\textwidth]{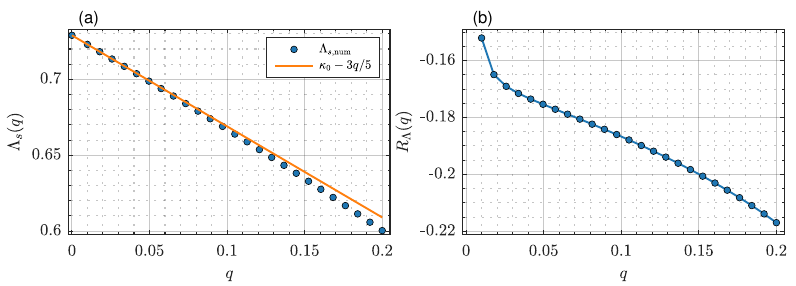}
\caption{Numerical check of the Laplace-domain Dirichlet--to--Neumann symbol. All quantities shown are dimensionless. The horizontal coordinate is \(q\), defined in \eqref{eq:q_symbol_validation}. (\emph{a}) Scaled finite-radius DtN symbol \(\Lambda_{\mathrm{s},\mathrm{num}}(q)=\Lambda_{\mathrm{num}}(p;\varepsilon)/p^{1/3}\), obtained from \eqref{eq:DtN_symbol_validation_ODE}--\eqref{eq:DtN_symbol_validation_BC}, compared with the first-order prediction \(\kappa_0-(3/5)q\). The marker at \(q=0\) denotes the planar Airy value \(\Lambda_{\mathrm{s}}(0)=\kappa_0\). (\emph{b}) Scaled residual \(R_\Lambda(q)\), defined in \eqref{eq:DtN_symbol_residual}.}
\label{fig:DtN_symbol_validation}
\end{figure}

Figure~\ref{fig:DtN_symbol_validation} compares
\(\Lambda_{\mathrm{s},\mathrm{num}}(q)\), obtained from
\eqref{eq:DtN_symbol_validation_ODE}--\eqref{eq:DtN_symbol_validation_BC},
with the first-order prediction \eqref{eq:scaled_symbol_prediction}.
Panel (a) shows the scaled finite-radius symbol as a function of the single
Laplace-domain curvature parameter \(q\). The marker at \(q=0\) denotes the
planar Airy limit, whereas the markers at \(q>0\) are finite-radius numerical
solutions. Panel (b) plots the scaled residual
\begin{equation}
  R_\Lambda(q)
  =
  \frac{
    \Lambda_{\mathrm{s},\mathrm{num}}(q)
    -\kappa_0
    +(3/5)q
  }{q^2},
  \qquad q>0.
  \label{eq:DtN_symbol_residual}
\end{equation}
The residual is not expected to vanish as \(q\to0\). Rather, if \(\Lambda_{\mathrm{s}}(q)=\kappa_0-(3/5)q+C_2q^2+O(q^3)\), then \(R_\Lambda(q)\to C_2\). Its boundedness confirms the \(O(q^2)\) remainder in the scaled symbol, equivalently the \(O(\varepsilon^2p^{-1/3})\) remainder in \eqref{eq:Lambda_asymp}. This calculation verifies the first-order coefficient in the Laplace-domain DtN symbol before any use is made of a particular wall condition.

\subsection{Range of validity}
\label{sec:results_validity}

The numerical comparisons support the ordering anticipated by the DtN symbol. For the uniformly active wall, the corrected model is accurate as long as the local curvature parameter \(\chi=\delta(z)/R=\varepsilon Z^{1/3}\) remains small. In this case curvature appears mainly as a local offset to the transfer coefficient. For the finite active-band wake, the same local correction enters a nonlocal closure and produces a slower-decaying contribution. The wake is therefore more sensitive to curvature in relative terms, even though the same small parameter controls the formal expansion.

The theory should consequently be interpreted as a weak-curvature expansion within the parabolic Graetz--L\'ev\^eque limit, not as a finite-curvature correlation and not as a full axial-diffusion model. Its useful range is set by two small parameters. The parabolic reduction of the full scalar advection--diffusion equation requires
\begin{equation}
  \beta=\frac{\delta_{\mathrm{b}}}{\ell_*}\ll1,
  \label{eq:validity_beta}
\end{equation}
whereas the local curvature expansion requires
\begin{equation}
  \chi(Z)=\frac{\delta(z)}{R}=\varepsilon Z^{1/3}\ll1.
  \label{eq:validity_chi}
\end{equation}
For reference, if \(\lambda=\ell_*/R\) and \(Pe_R=U_{\mathrm{b}}R/D\), then
\begin{equation}
  \varepsilon=\left(\frac{\lambda}{4Pe_R}\right)^{1/3},
  \qquad
  \beta=\left(\frac{1}{4Pe_R\lambda^2}\right)^{1/3}
  =
  \frac{\varepsilon}{\lambda}.
  \label{eq:validity_eps_beta_pe}
\end{equation}
Thus increasing \(Pe_R\) decreases both \(\varepsilon\) and \(\beta\), while the ratio \(\varepsilon/\beta=\lambda\) is fixed by the chosen streamwise length scale. Within the regime \(\beta\ll1\) and \(\chi(Z)\ll1\), the first-order curvature correction captures both the constant Sherwood-number shift for a uniformly active wall and the curvature-induced enhancement of the downstream wake under mixed wall conditions. The immediate neighbourhoods of the leading edge and the active--insulated switching point remain outside the parabolic approximation because axial diffusion regularises the corresponding edge layers. {The effect of retaining axial diffusion on the downstream wake is assessed for a representative case in \S\ref{sec:results_decay}.}

{An order-of-magnitude estimate of the switching region follows from the new near-wall adjustment layer generated by the change to zero wall flux. At a distance \(s=z-\ell\ll\ell\) downstream, shear advection and wall-normal diffusion give its thickness \(h_s\sim(Ds/S_{\mathrm{w}})^{1/3}\), distinct from the scalar layer already present at the end of the active band. The ratio of axial to wall-normal diffusion for this adjustment is therefore \((h_s/s)^2\sim\beta^2(Z-1)^{-4/3}\). Axial diffusion becomes comparable when \(s=O(\sqrt{D/S_{\mathrm{w}}})\), or \(Z-1=O(\beta^{3/2})\), so the local parabolic description requires \(Z-1\gg\beta^{3/2}\). For \(\beta=0.05\), this scale is approximately \(0.011\), placing \(Z=1.001\) within the estimated switching region and the comparison points \(Z=1.5\), \(2\) and \(5\) well beyond it. This estimate identifies the local breakdown scale of the parabolic approximation; it does not quantify the resulting change in the downstream wake.}

\par\begingroup\color{black}
An illustrative choice \(Pe_R=10^6\) and \(\lambda=1\) gives \(\varepsilon=\beta\simeq0.00630\). The first-order formula \eqref{eq:single_electrode_flux_dimless} then predicts a local flux approximately \(0.56\,\%\) below the planar value at the midpoint of the active band, \(Z=0.5\). In the wake, \eqref{eq:g0_beta} and \eqref{eq:g1_fractional_integral} give relative increases in wall deficit, \(\varepsilon G_1/G_0\), of approximately \(0.53\,\%\), \(0.70\,\%\) and \(1.23\,\%\) at \(Z=1.5\), \(2\) and \(5\), respectively. These estimates illustrate the downstream growth of the relative curvature effect even when its influence on active-band transfer is sub-percent.
\par\endgroup

\section{Conclusion}
\label{sec:conclusion}

For a finite active annular wall band in pipe flow, finite radius has two linked transport consequences: it reduces transfer over the active band and increases the residual scalar deficit on the downstream insulated wall. We have derived a weak-curvature correction to the Graetz--L\'ev\^eque wall-flux law for passive-scalar transport in fully developed pipe flow. The analysis was carried out in the parabolic high-P\'eclet-number boundary-layer limit \(\beta=\delta_{\mathrm{b}}/\ell_*\ll1\), where axial diffusion is asymptotically smaller than wall-normal diffusion away from axial-diffusion edge layers. Within this limit, finite pipe radius affects the near-wall transport through two first-order mechanisms: the cylindrical metric correction in the diffusion operator and the quadratic finite-radius correction to the near-wall Poiseuille velocity.

The central result is the curvature-corrected Dirichlet--to--Neumann map relating the wall scalar trace \(G\) to the wall flux \(J\). In Laplace space, weak curvature shifts the planar L\'ev\^eque symbol \(\kappa_0p^{1/3}\) by the additive local term \(-3\varepsilon/5\), with the next correction ordered by \(\varepsilon^2p^{-1/3}\). Thus the planar contribution is fractional and nonlocal in the streamwise coordinate, whereas the first curvature correction is local in physical space. The coefficient \(-3/5\) contains one contribution from the cylindrical diffusion metric and another from the finite-radius correction to the velocity profile. The ordering of the remainder shows that the relevant local small parameter is the scalar-layer thickness relative to the pipe radius, \(\delta(z)/R\sim\varepsilon Z^{1/3}\), not the reference curvature parameter \(\varepsilon\) alone.

For a uniformly active wall, the curvature-corrected map recovers the classical L\'ev\^eque entrance exponent and adds a constant negative correction to the wall flux. In the local, diameter-based Sherwood expansion used here, this appears as the constant term \(-6/5\). This uniform-wall result is a consistency check against the extended L\'ev\^eque expansion for a circular tube. The main new consequence appears for mixed wall conditions. For a finite annular active band followed by an insulated wall, the zero-flux condition in the wake is imposed through the nonlocal Dirichlet--to--Neumann map. The planar wake trace decays as \(G_0\sim Z^{-2/3}\), whereas the first curvature correction decays as \(G_1\sim Z^{-1/3}\). For \(\varepsilon=0.1\), this correction reduces the planar wake error from \(8.7\,\%\), \(11.5\,\%\) and \(20.2\,\%\) to \(1.1\,\%\), \(1.7\,\%\) and \(4.7\,\%\) at \(Z=1.5\), \(2\) and \(5\), respectively. A local \(O(\varepsilon)\) correction to the wall-flux law is therefore converted by the mixed Dirichlet--Neumann closure into a more slowly decaying downstream wall-deficit response.

The asymptotic predictions were tested against direct numerical solutions of the finite-radius cylindrical boundary-layer problem. A separate Laplace-domain calculation verifies the first-order Dirichlet--to--Neumann symbol coefficient before imposing any particular wall condition. Marching solutions of the finite-radius parabolic problem then confirm the uniform-wall flux correction, the Sherwood-number shift and the curvature-induced wake response. These tests support the weak-curvature expansion within the parabolic Graetz--L\'ev\^eque model. They do not constitute a full axial-diffusion validation; in the full scalar problem, abrupt changes in wall condition would be regularised by axial-diffusion edge layers. A targeted full-equation calculation at \(\varepsilon=0.1\) and \(\beta=0.05\) gives wake changes of \(0.40\,\%\) and \(0.21\,\%\) at the interior positions \(Z=1.5\) and \(2\), respectively, small relative to the retained curvature correction for the boundary conditions tested.

The resulting picture is therefore clear. Weak curvature produces a local transfer-coefficient offset for a uniformly active wall, but the same local offset has a longer-ranged effect when the wall condition is mixed. The formulation provides a compact operator-level description of finite-radius effects in Graetz--L\'ev\^eque transport and can be extended to higher-order curvature corrections, more general axially patterned active--passive bands, and unsteady or frequency-domain transfer problems. Its present range of validity is the regime \(\beta\ll1\) and \(\delta(z)/R\ll1\); outside that regime, axial-diffusion edge layers or finite-curvature effects must be treated separately. Within this range, the result quantifies the finite-radius error in planar predictions of scalar recovery behind an active band, providing the wall response needed to model transport between successive active regions. For wall-transfer measurements, this helps distinguish changes in sensor response caused by finite radius and upstream scalar depletion from changes caused by wall shear.

\section*{Acknowledgements}
OpenAI ChatGPT (GPT-5.5 Pro, web interface, May--June 2026) was used for language editing and manuscript preparation assistance; the authors reviewed and approved the final manuscript.

\section*{Funding}
This publication was supported by the European Union under the Operational Programme Johannes Amos Comenius (OP JAC), project MATBIOMED -- Materials and Technologies for Bioapplications and Medicine, project registration no. CZ.02.01.01/00/23\_021/0008398, and by the Internal Grant Agency of Jan Evangelista Purkyn\v{e} University in \'{U}st\'{i} nad Labem (project no. UJEP-IGA-2026-53-002-2).

\section*{Declaration of interests}
The authors report no conflict of interest.

\appendix

\section{Numerical implementation and auxiliary checks}
\label{sec:appendix_benchmark}

This appendix gives the numerical details behind the \(\beta=0\) finite-radius boundary-layer benchmarks used in \S\ref{sec:results}. The separate full-equation check of axial diffusion is described in \S\ref{sec:results_decay}. The purpose of the numerical work is therefore to test the weak-curvature expansion against the finite-radius boundary-layer model from which it is derived.

\subsection{Finite-radius boundary-layer marching problem}
\label{sec:appendix_BL_marching}

The finite-radius benchmark solves
\begin{equation}
  a(X)\frac{\partial\Theta}{\partial Z}
  =
  \mathcal{L}_\varepsilon\Theta,
  \qquad
  a(X)=X\left(1-\frac{\varepsilon X}{2}\right),
  \qquad
  0<X<\varepsilon^{-1},
  \label{eq:appendix_BL_problem}
\end{equation}
where
\begin{equation}
  \mathcal{L}_\varepsilon\Theta
  =
  \Theta_{XX}
  -
  \frac{\varepsilon}{1-\varepsilon X}\Theta_X
  =
  \frac{1}{1-\varepsilon X}
  \frac{\partial}{\partial X}
  \left[
    (1-\varepsilon X)\Theta_X
  \right].
  \label{eq:appendix_conservative_operator}
\end{equation}
The conservative form in \eqref{eq:appendix_conservative_operator} was used for the wall-normal discretisation. On a cell-centred mesh, the discrete operator is obtained by evaluating the flux \((1-\varepsilon X)\Theta_X\) at cell faces. The outermost face coincides with the pipe axis, where \(1-\varepsilon X=0\); the finite-volume flux through this face therefore vanishes, which is the discrete form of the axisymmetry condition. At the wall, either the Dirichlet condition \(\Theta(0,Z)=0\) or the Neumann condition \(\Theta_X(0,Z)=0\) is imposed, depending on whether the wall is active or insulated.

Marching in \(Z\) was performed with a fully implicit first-order step,
\begin{equation}
  a_i\frac{\Theta_i^{n+1}-\Theta_i^n}{\Delta Z_n}
  =
  \left(\mathcal{L}_\varepsilon\Theta^{n+1}\right)_i,
  \label{eq:appendix_implicit_step}
\end{equation}
which gives a tridiagonal linear system at each axial step. The inlet condition is \(\Theta(X,0)=1\). For the uniformly active wall, the wall is Dirichlet for all \(Z>0\). For the finite annular active band, the wall is Dirichlet on \(0<Z<1\) and Neumann for \(Z>1\). The wall flux \(J=\Theta_X(0,Z)\) was evaluated using a one-sided wall approximation consistent with the wall-normal discretisation.

The computations used in the main figures employed \(N_X=1200\) cell-centred wall-normal points. The uniformly active-wall calculations used 1200 logarithmically spaced axial points on \(10^{-3}\le Z\le1\). The finite active-band calculations used 1200 logarithmically spaced points on \(10^{-3}\le Z\le1\) and 1800 linearly spaced points on \(1<Z\le5\). \par\begingroup\color{black}
The resolution of the numerical benchmark was quantified with an additional independent conservative cell-centred calculation for the uniformly active wall at \(\varepsilon=0.1\). The finite-radius flux was computed on \(N_X=400,800,1600,3200\) wall-normal cells and with 600, 1200, 2400 and 4800 logarithmically spaced axial output points, with a separately resolved logarithmic entrance mesh. Table~\ref{tab:uniform_resolution} summarises the flux and its separate resolution sensitivities. The fully implicit axial step showed the expected first-order behaviour: at \(Z=0.02\), \(0.1\) and \(1\), increasing the axial resolution from 2400 to 4800 changed the flux by \(0.0167\,\%\), \(0.0169\,\%\) and \(0.0181\,\%\), respectively. At 4800 axial points, increasing \(N_X\) from 1600 to 3200 changed these fluxes by \(0.0104\,\%\), \(0.0104\,\%\) and \(0.0104\,\%\). Changing the wall derivative from a two-cell to a four-cell one-sided expression on the finest radial mesh changed the same values by approximately \(0.0027\,\%\). These sensitivities are reported separately; they are not combined as a rigorous global error bound. The finite-radius flux remains below the planar prediction on every grid. Thus the numerical check supports the observed improvement of the curvature-corrected approximation, while its smallest quoted discrepancy, approximately \(0.03\,\%\), is interpreted as an approximate difference from the benchmark rather than as a precision-certified value. 

\begin{table}\begin{center}\color{black}\captionsetup{font=footnotesize,labelfont={footnotesize,color=black},textfont={footnotesize,color=black}}\begin{tabular}{c c c c c} \(Z\) & \(J\) & \(\Delta_Z\) (\%) & \(\Delta_X\) (\%) & \(\Delta_J\) (\%) \\[3pt] \(0.02\) & \(1.922955\) & \(0.0167\) & \(0.0104\) & \(0.0027\) \\ \(0.10\) & \(1.099035\) & \(0.0169\) & \(0.0104\) & \(0.0027\) \\ \(1.00\) & \(0.476317\) & \(0.0181\) & \(0.0104\) & \(0.0027\) \\ \end{tabular}\caption{Resolution sensitivity for the uniformly active wall at \(\varepsilon=0.1\). The reference flux \(J\) uses \(N_X=3200\), \(N_Z=4800\). \(\Delta_Z\): change from \(N_Z=2400\) to \(4800\) at \(N_X=1600\); \(\Delta_X\): change from \(N_X=1600\) to \(3200\) at \(N_Z=4800\); \(\Delta_J\): change of the wall-derivative stencil on the finest mesh. All changes are relative percentages.}\label{tab:uniform_resolution}\end{center}\end{table}
\par\endgroup

\par\begingroup\color{black}
An analogous independent resolution check was performed for the finite active band at \(\varepsilon=0.1\), with the wall condition changed from Dirichlet to zero flux exactly at \(Z=1\). Table~\ref{tab:wake_resolution} summarises the downstream wall deficit and its separate resolution sensitivities. At \(Z=1.5\), \(2\) and \(5\), increasing the number of axial points in each of the active and wake intervals from 2400 to 4800 at \(N_X=800\) changed the wake trace by \(0.0157\,\%\), \(0.0277\,\%\) and \(0.0519\,\%\), respectively. At the latter axial resolution, increasing \(N_X\) from 800 to 1600 changed the same values by \(5.5\times10^{-4}\,\%\), \(5.1\times10^{-4}\,\%\) and \(4.5\times10^{-4}\,\%\). Replacing the quadratic zero-flux wall extrapolation by a fourth-degree polynomial extrapolation changed the trace by less than \(6.3\times10^{-6}\,\%\) at these points. The discrete integral of the scalar deficit in the insulated wake was conserved to within \(1.2\times10^{-11}\) relative. On \(1.2\leq Z\leq5\), the finest independent trace differs from the benchmark {used in figures~\ref{fig:wake_validation} and~\ref{fig:wake_decay}} by at most \(0.0692\,\%\). The immediate switching neighbourhood is reported separately: at \(Z=1.001\), the difference from the {benchmark} is \(0.843\,\%\), whereas the final axial change is only \(1.8\times10^{-4}\,\%\). This local discrepancy is therefore not used to quantify the resolved downstream wake or the error of the curvature expansion.

\begin{table}\begin{center}\color{black}\captionsetup{font=footnotesize,labelfont={footnotesize,color=black},textfont={footnotesize,color=black}}\begin{tabular}{c c c c c} \(Z\) & \(G\) & \(\Delta_Z\) (\%) & \(\Delta_X\) (\%) & \(\Delta_G\) (\%) \\[3pt] \(1.5\) & \(0.447196\) & \(0.0157\) & \(5.5\times10^{-4}\) & \(<6.3\times10^{-6}\) \\ \(2.0\) & \(0.349126\) & \(0.0277\) & \(5.1\times10^{-4}\) & \(<6.3\times10^{-6}\) \\ \(5.0\) & \(0.18798{5}\) & \(0.0519\) & \(4.5\times10^{-4}\) & \(<6.3\times10^{-6}\) \\ \end{tabular}\caption{Resolution sensitivity for the finite-band wake at \(\varepsilon=0.1\). The reference deficit \(G\) uses \(N_X=1600\), \(N_Z=4800\). \(\Delta_Z\): change from \(N_Z=2400\) to \(4800\) at \(N_X=800\); \(\Delta_X\): change from \(N_X=800\) to \(1600\) at \(N_Z=4800\); \(\Delta_G\): change of the zero-flux wall extrapolation on the finest mesh. All changes are relative percentages.}\label{tab:wake_resolution}\end{center}\end{table}
\par\endgroup

\subsection{Numerical solution for the Laplace-domain DtN symbol}
\label{sec:appendix_DtN_symbol_solver}

The Laplace-domain symbol check in \S\ref{sec:results_DtN_symbol} was obtained independently of the \(Z\)-marching calculation. After introducing \(y=p^{1/3}X\) and \(q=\varepsilon p^{-1/3}\), the finite-radius Laplace-domain DtN problem for \(q>0\) is
\begin{equation}
  \frac{\mathrm d}{\mathrm dy}
  \left[
    (1-qy)\frac{\mathrm d\Psi}{\mathrm dy}
  \right]
  -
  y\left(1-\frac{qy}{2}\right)(1-qy)\Psi
  =0,
  \qquad
  0<y<q^{-1},
  \label{eq:appendix_DtN_symbol_ode}
\end{equation}
with
\begin{equation}
  \Psi(0)=1,
  \qquad
  \Psi_y(q^{-1})=0.
  \label{eq:appendix_DtN_symbol_bc}
\end{equation}
The scaled numerical symbol is
\begin{equation}
  \Lambda_{\mathrm{s},\mathrm{num}}(q)
  =
  \frac{\Lambda_{\mathrm{num}}(p;\varepsilon)}{p^{1/3}}
  =
  -\Psi_y(0).
  \label{eq:appendix_scaled_symbol}
\end{equation}
Thus the computation produces the DtN multiplier directly, rather than a physical-space concentration field.
Equation \eqref{eq:appendix_DtN_symbol_ode} was discretised directly in conservative form on a uniform grid in \(y\), with the wall value imposed at \(y=0\) and the axis condition imposed at \(y=q^{-1}\). A fourth-order one-sided derivative was used to evaluate \(\Psi_y(0)\). The case \(q=0\) was treated separately as the planar half-line Airy problem, giving \(\Lambda_{\mathrm{s}}(0)=\kappa_0\). The calculation in figure~\ref{fig:DtN_symbol_validation} used \(0\le q\le0.2\); the plotted scaled residual remained \(O(1)\), confirming the \(O(q^2)\) remainder in the scaled symbol.

\subsection{Independent radial-coordinate implementation check}
\label{sec:appendix_rho_solver}

As an additional implementation check, the same finite-radius parabolic Graetz problem was also solved in the original radial coordinate \(\rho=r/R\). This is not a separate physical model: under the change of variables \(\rho=1-\varepsilon X\), it is algebraically equivalent to \eqref{eq:appendix_BL_problem}. Its value is that it provides an independent coordinate implementation of the finite-radius benchmark.

In the radial coordinate, the equation is
\begin{equation}
  \frac{1-\rho^2}{2\varepsilon^3}\frac{\partial\Theta}{\partial Z}
  =
  \frac{1}{\rho}\frac{\partial}{\partial\rho}
  \left(
    \rho\frac{\partial\Theta}{\partial\rho}
  \right),
  \qquad
  0<\rho<1.
  \label{eq:rho_coordinate_graetz}
\end{equation}
The axis condition is \(\Theta_\rho(0,Z)=0\). At the wall, either \(\Theta(1,Z)=0\) or \(\Theta_\rho(1,Z)=0\) is imposed. The dimensionless flux used in the main text is recovered from
\begin{equation}
  J(Z)
  =
  -\varepsilon\left.\frac{\partial\Theta}{\partial\rho}\right|_{\rho=1}.
  \label{eq:rho_flux_relation}
\end{equation}
The relative differences shown below are defined with the \(X\)-coordinate calculation as the reference:
\begin{equation}
  \Delta_J(Z)
  =
  \frac{|J_\rho(Z)-J_X(Z)|}{|J_X(Z)|},
  \qquad
  \Delta_G(Z)
  =
  \frac{|G_\rho(Z)-G_X(Z)|}{|G_X(Z)|},
  \label{eq:rho_solver_relative_differences}
\end{equation}
where the subscripts \(\rho\) and \(X\) denote the radial-coordinate and wall-normal-coordinate implementations, respectively.
Equation \eqref{eq:rho_coordinate_graetz} was discretised by a conservative cell-centred finite-volume method in \(\rho\) and advanced implicitly in \(Z\), independently of the \(X\)-coordinate implementation.

\begin{figure}
\centering
\includegraphics[width=\textwidth]{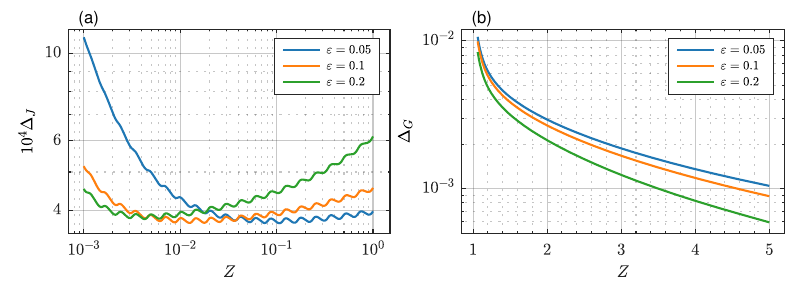}
\caption{Independent radial-coordinate check of the finite-radius boundary-layer benchmark. (\emph{a}) Relative difference \(\Delta_J\), defined in \eqref{eq:rho_solver_relative_differences}, for the uniformly active wall. (\emph{b}) Relative difference \(\Delta_G\), defined in \eqref{eq:rho_solver_relative_differences}, for the wake behind a finite annular active band, excluding the immediate switching neighbourhood. The radial-coordinate calculation solves \eqref{eq:rho_coordinate_graetz}; the reference solution is the \(X\)-coordinate benchmark used in \S\ref{sec:results}.}
\label{fig:appendix_rho_solver_check}
\end{figure}

Figure~\ref{fig:appendix_rho_solver_check} compares the radial-coordinate implementation with the \(X\)-coordinate benchmark used in the main text. For the uniformly active wall the relative difference in the flux is below approximately \(1.1\times10^{-3}\) over the plotted interval. In the wake calculation the largest differences occur close to the switching point, where the wall condition changes; away from this immediate neighbourhood the relative difference decreases downstream and is typically of order \(10^{-3}\). This check confirms that the numerical comparisons in \S\ref{sec:results} are not an artefact of a particular coordinate implementation.

\section{Closed form of the first wake correction}
\label{sec:appendix_g1_closed}

The main text uses the fractional-integral representation \eqref{eq:g1_fractional_integral} because it displays the operator structure of the curvature correction. For completeness, this appendix gives an equivalent closed form in terms of lower incomplete beta functions. We use \(B_x(a,b)\) for the lower incomplete beta function and \(B(a,b)=B_1(a,b)\) for the complete beta function.

The planar wake trace is
\begin{equation}
  G_0(S)
  =
  \frac{
    B_{1/S}\!\left(\tfrac23,\tfrac13\right)
  }{
    B\!\left(\tfrac23,\tfrac13\right)
  }
  =
  \frac{1}{B\!\left(\tfrac23,\tfrac13\right)}
  \int_0^{1/S} t^{-1/3}(1-t)^{-2/3}\,\mathrm{d}t,
  \qquad S\ge1.
  \label{eq:g0_euler_repr_appendix}
\end{equation}
Substitution into \eqref{eq:g1_fractional_integral} gives
\begin{equation}
  G_1(Z)
  =
  \frac{3}{5\kappa_0\Gamma(1/3)} I(Z),
  \qquad
  I(Z)=\int_1^Z (Z-S)^{-2/3}G_0(S)\,\mathrm{d}S.
  \label{eq:g1_appendix_I_def}
\end{equation}
For \(Z>1\), interchanging the order of integration in \eqref{eq:g1_appendix_I_def} yields
\begin{equation}
  I(Z)
  =
  3(Z-1)^{1/3}
  -
  \frac{3}{B\!\left(\tfrac23,\tfrac13\right)}
  \int_{1/Z}^{1}
    t^{-2/3}(1-t)^{-2/3}(Zt-1)^{1/3}
  \,\mathrm{d}t.
  \label{eq:I_reduced_appendix}
\end{equation}
The remaining integral can be expressed in incomplete-beta form, giving
\begin{equation}
  I(Z)
  =
  3(Z-1)^{1/3}
  -
  2\,\frac{B\!\left(\tfrac43,\tfrac13\right)}{B\!\left(\tfrac23,\tfrac13\right)}
  Z^{1/3}
  B_{(Z-1)/Z}\!\left(\tfrac23,\tfrac23\right),
  \qquad Z>1.
  \label{eq:g1_fractional_integral_closed_appendix}
\end{equation}
Consequently,
\begin{equation}
  G_1(Z)
  =
  \frac{3}{5\kappa_0\Gamma(1/3)}
  \left[
    3(Z-1)^{1/3}
    -
    2\,\frac{B\!\left(\tfrac43,\tfrac13\right)}{B\!\left(\tfrac23,\tfrac13\right)}
    Z^{1/3}
    B_{(Z-1)/Z}\!\left(\tfrac23,\tfrac23\right)
  \right],
  \qquad Z>1.
  \label{eq:g1_closed_form_appendix}
\end{equation}
This expression is algebraically equivalent to the convolution form used in the main text, but the latter is preferable for interpreting the wake correction as a fractional integral of the planar wake.

\section{Conversion to the local diameter-based Sherwood expansion}
\label{sec:appendix_extended_leveque}

This appendix records the conversion between the shear-based variables used in the analysis and the local, inlet/outer-referenced, diameter-based Sherwood--Graetz notation used here. The purpose is to make the normalisation of the constant curvature correction unambiguous before comparison with extended L\'ev\^eque results for circular tubes.

The local transfer coefficient for a uniformly active wall is
\begin{equation}
  k(z)
  =
  \frac{3^{1/3}}{\Gamma(1/3)}
  \left(\frac{D^2S_{\mathrm{w}}}{z}\right)^{1/3}
  -
  \frac{3}{5}\frac{D}{R}
  +
  O\!\left(
    \frac{D}{R}\frac{\delta(z)}{R}
  \right),
  \qquad
  \delta(z)=\left(\frac{Dz}{S_{\mathrm{w}}}\right)^{1/3}.
  \label{eq:appendix_k_uniform}
\end{equation}
Using \(U_{\mathrm{b}}=S_{\mathrm{w}}R/4\), the diameter-based Graetz number is
\begin{equation}
  \mathrm{Gz}_D
  =
  \frac{U_{\mathrm{b}}(2R)^2}{Dz}
  =
  \frac{S_{\mathrm{w}}R^3}{Dz}
  =
  \left(\frac{R}{\delta(z)}\right)^3.
  \label{eq:appendix_Gz_conversion}
\end{equation}
The local Sherwood number used here is based on the diameter \(2R\) and on the inlet/outer scalar difference, so that
\begin{equation}
  \mathrm{Sh}_D
  =
  \frac{2Rk(z)}{D}.
  \label{eq:appendix_Sh_def}
\end{equation}
Combining \eqref{eq:appendix_k_uniform}--\eqref{eq:appendix_Sh_def} gives
\begin{equation}
  \mathrm{Sh}_D
  =
  \frac{2\,3^{1/3}}{\Gamma(1/3)}\mathrm{Gz}_D^{1/3}
  -
  \frac{6}{5}
  +
  O\!\left(\mathrm{Gz}_D^{-1/3}\right).
  \label{eq:appendix_Sh_Gz_final}
\end{equation}
Numerically,
\begin{equation}
  \frac{2\,3^{1/3}}{\Gamma(1/3)}=1.076732\ldots,
  \qquad
  \frac{6}{5}=1.2.
  \label{eq:appendix_Sh_coefficients}
\end{equation}
Thus, in the present convention for the local, diameter-based Sherwood number and Graetz number,
\begin{equation}
  \mathrm{Sh}_D
  =
  1.076732\,\mathrm{Gz}_D^{1/3}
  -
  1.2
  +
  O\!\left(\mathrm{Gz}_D^{-1/3}\right).
  \label{eq:appendix_Sh_numeric}
\end{equation}
This is the form to be compared with extended L\'ev\^eque expansions for a circular tube after matching the Sherwood and Graetz conventions \citep{WorsoeSch1967,Newman1969}. Different expansion coefficients are reported in the literature when the Sherwood number is averaged over the active length, when a bulk rather than inlet/outer scalar reference is used, when the hydraulic radius rather than the diameter is used as reference length, or when a different definition of the Graetz number is adopted. Equation \eqref{eq:appendix_Sh_Gz_final} fixes the convention used in the present paper. For the uniform active wall considered here, using the flow-weighted bulk scalar difference instead of the inlet difference leaves the constant term \(-6/5\) unchanged; the distinction first enters at order \(\mathrm{Gz}_D^{-1/3}\).

In Newman's extended L\'ev\^eque solution \citep[equations (18)--(19)]{Newman1967Report}, the first wall-gradient correction is \(-0.6\), and the constant term in the length-averaged, inlet-referenced Nusselt number is \(-1.2\). The correction \(-3D/(5R)\) in \eqref{eq:appendix_k_uniform} gives the same constant \(-1.2\) in the local diameter-based Sherwood number. Averaging this constant contribution along the active length leaves it unchanged, although it changes the coefficient of the leading L\'ev\^eque term. Thus the local and length-averaged expansions reproduce the same first finite-radius offset when both use the inlet scalar difference and the pipe diameter.

\par\begingroup\color{black}
\renewcommand{\theHsection}{appendix.\Alph{section}}
\renewcommand{\theHequation}{appendix.\Alph{section}.\arabic{equation}}
\section{Similarity derivation of the far-wake curvature correction}
\label{sec:appendix_wake_similarity}

Conservation separates the effect of reduced active-band uptake from
the redistribution that produces the slower wall-deficit correction.
Let \(\eta=XZ^{-1/3}\) and \(F(\eta)=\exp(-\eta^3/9)\). The leading
planar wake is \(\vartheta_0\sim a_0Z^{-2/3}F(\eta)\), where
\begin{equation}
I_F=\int_0^\infty\eta F(\eta)\,\mathrm d\eta
    =3^{1/3}\Gamma(2/3),\qquad
\mathcal M_0=\frac{3^{4/3}}{2\Gamma(1/3)},\qquad
a_0=\frac{\mathcal M_0}{I_F}=\frac{3\sqrt3}{4\pi}.
\label{eq:wake_similarity_planar}
\end{equation}
Here \(\mathcal M_0\) follows by integrating the planar active-band
flux over \(0<Z<1\). Integrating its first correction gives
\(\mathcal M_\varepsilon=\mathcal M_0-3\varepsilon/5+O(\varepsilon^2)\).
At first order the far-wake expansion is
\begin{equation}
\vartheta_1
=Z^{-1/3}g(\eta)+a_1Z^{-2/3}F(\eta)+o(Z^{-2/3}).
\label{eq:wake_similarity_split}
\end{equation}
Inserting this expansion into \eqref{eq:wake_deficit_balance}, with
the upper limit extended to infinity in the thin-layer limit, gives
\begin{equation}
\mathcal M_1
=Z^{1/3}\left[
  \int_0^\infty\eta g(\eta)\,\mathrm d\eta
  -\frac{3a_0}{2}\int_0^\infty\eta^2F(\eta)\,\mathrm d\eta
 \right]+a_1 I_F+o(1).
\label{eq:wake_similarity_balance}
\end{equation}
The flux is independent of \(Z\), and \(\int_0^\infty\eta^2F\,\mathrm
d\eta=3\). Therefore
\begin{equation}
\int_0^\infty\eta g(\eta)\,\mathrm d\eta=\frac{9a_0}{2},
\qquad
a_1 I_F=-\frac35.
\label{eq:wake_similarity_moments}
\end{equation}
The growing contribution to the integral balance is cancelled by
redistribution; the constant first-order uptake change fixes \(a_1\).

The wall value of \(g\), unlike its integral moment, also requires the
transverse scalar equation. Expanding \eqref{eq:graetz_BL_XZ} gives
\begin{equation}
X\partial_Z\vartheta_1-\partial_{XX}\vartheta_1
=\frac{X^2}{2}\partial_Z\vartheta_0-\partial_X\vartheta_0.
\label{eq:wake_similarity_first_order}
\end{equation}
Substitution of the similarity forms yields
\begin{equation}
g''+\frac{\eta}{3}(g+\eta g')
=-\frac{a_0}{18}\eta^5\exp(-\eta^3/9),
\qquad g'(0)=0,\qquad g(\eta)\to0,
\qquad \int_0^\infty\eta|g(\eta)|\,\mathrm d\eta<\infty.
\label{eq:wake_similarity_ode}
\end{equation}
Primes denote differentiation with respect to \(\eta\).
The finite-moment condition is essential: wall symmetry and decay
alone admit an additional homogeneous solution with a \(1/\eta\)
tail, which cannot carry a finite advective deficit flux.

An explicit finite-moment solution is
\begin{equation}
\begin{aligned}
g(\eta)&=a_0e^{-t}\left[
 \frac{\eta^4}{30}+\frac{3\eta}{5}
 +c\,U\!\left(\frac13,\frac23,t\right)\right],
\qquad t=\frac{\eta^3}{9},\\
c&=\frac{9^{1/3}\Gamma(1/3)}{5\Gamma(2/3)}.
\end{aligned}
\label{eq:wake_similarity_solution}
\end{equation}
Here \(U\) is Tricomi's confluent hypergeometric function; for the
parameters above it has the integral representation
\[
U\!\left(\frac13,\frac23,t\right)
=\frac{1}{\Gamma(1/3)}
\int_0^\infty e^{-ts}s^{-2/3}(1+s)^{-2/3}\,\mathrm ds.
\]
Substitution verifies \eqref{eq:wake_similarity_ode}; the coefficient
\(c\) enforces \(g'(0)=0\), and the exponential factor ensures a finite
moment. Since \(U(1/3,2/3,0)=\Gamma(1/3)/\Gamma(2/3)\), the wall
amplitude is
\begin{equation}
g(0)=\frac{a_0\,9^{1/3}\Gamma(1/3)^2}{5\Gamma(2/3)^2}
     =\frac{9}{10\kappa_0\Gamma(2/3)^2}.
\label{eq:wake_similarity_amplitude}
\end{equation}
This recovers \eqref{eq:wake_asymptotics} from the transverse
similarity problem rather than from fractional integration of the
wall trace. It also shows \(g>0\), consistently with
\eqref{eq:g1_fractional_integral}. The expansion describes
\(1\ll Z\ll\varepsilon^{-3}\), not the fully mixed finite-radius limit.
\par\endgroup

\bibliographystyle{jfm}
\begingroup\let\jfmOriginalBibitem\bibitem\renewcommand{\bibitem}[2][]{\jfmOriginalBibitem[#1]{#2}\def\jfmRefKey{#2}\def\jfmThermalRef{BellhouseSchultz1966}\def\jfmElectroRef{MitchellHanratty1966}\def\jfmNewmanReportRef{Newman1967Report}\def\jfmHuchetRef{Huchet2011}\def\jfmSirkarRef{SirkarHanratty1970}\ifx\jfmRefKey\jfmSirkarRef\color{black}\else\ifx\jfmRefKey\jfmHuchetRef\color{black}\else\ifx\jfmRefKey\jfmNewmanReportRef\color{black}\else\ifx\jfmRefKey\jfmThermalRef\color{black}\else\ifx\jfmRefKey\jfmElectroRef\color{black}\else\color{black}\fi\fi\fi\fi\fi}\bibliography{curvature_graetz}\endgroup

\end{document}

%% file: arxiv-layout.tex
\usepackage{etoolbox}
\pdfpagewidth=\paperwidth
\pdfpageheight=\paperheight
\makeatletter
\patchcmd{\@maketitle}{\vspace*{10\p@}\addvspace{4.6pc}}{\vspace*{0pt}}{}
  {\PackageError{arxiv-layout}{Cannot remove journal banner spacing}{}}
\patchcmd{\@maketitle}{(Received xx; revised xx; accepted xx)}
  {Author's accepted manuscript. Accepted for publication in \emph{Journal of Fluid Mechanics}.}{}
  {\PackageError{arxiv-layout}{Cannot replace journal history placeholders}{}}
\def\ps@arxiv{%
  \let\@mkboth\@gobbletwo
  \def\@oddhead{\hfil\itshape\@righttitle\hfil}%
  \def\@evenhead{\hfil\itshape\@lefttitle\hfil}%
  \def\@oddfoot{\hfil\normalfont\thepage\hfil}%
  \let\@evenfoot\@oddfoot
  \def\sectionmark##1{}\def\subsectionmark##1{}}
\def\ps@titlepage{\ps@arxiv\def\@oddhead{}\def\@evenhead{}}
\AtBeginDocument{\pagestyle{arxiv}}
\makeatother
\hypersetup{pdftitle={Curvature effects on Graetz-Leveque transport from finite wall bands in pipe flow},
  pdfauthor={Jaromir Havlica; Yaroslav Bazaikin; Vaclav Harrandt; Florian Huchet; Jaroslav Tihon},
  pdfsubject={Author's accepted manuscript for Journal of Fluid Mechanics},
  hypertexnames=false,colorlinks=false,pdfborder={0 0 0}}

%% file: main.bbl
\begin{thebibliography}{26}
\expandafter\ifx\csname natexlab\endcsname\relax\def\natexlab#1{#1}\fi
\def\au#1{#1} \def\ed#1{#1} \def\yr#1{#1}\def\at#1{#1}\def\jt#1{\textit{#1}}
  \def\bt#1{#1}\def\bvol#1{\textbf{#1}} \def\vol#1{#1} \def\pg#1{#1}
  \def\publ#1{#1}\def\arxiv#1{#1}\def\org#1{#1}\def\st#1{\textit{#1}}

\bibitem[Ackerberg {\em et~al.\/}(1978)Ackerberg, Patel \&
  Gupta]{AckerbergPatelGupta1978}
{\sc \au{Ackerberg, R.~C.}, \au{Patel, R.~D.} \& \au{Gupta, S.~K.}} \yr{1978}
  \at{The heat/mass transfer to a finite strip at small {P}éclet numbers}.
  \jt{Journal of Fluid Mechanics}  \bvol{86}~(1),  \pg{49--65}.

\bibitem[Bellhouse \& Schultz(1966)]{BellhouseSchultz1966}
{\sc \au{Bellhouse, B.~J.} \& \au{Schultz, D.~L.}} \yr{1966}  \at{Determination
  of mean and dynamic skin friction, separation and transition in low-speed
  flow with a thin-film heated element}.  \jt{Journal of Fluid Mechanics}
  \bvol{24}~(2),  \pg{379--400}.

\bibitem[Deen(2012)]{Deen2012}
{\sc \au{Deen, W.~M.}} \yr{2012} {\em Analysis of Transport Phenomena\/}, 2nd
  edn.  \publ{New York: Oxford University Press}.

\bibitem[Graetz(1885)]{Graetz1885}
{\sc \au{Graetz, L.}} \yr{1885}  \at{{Ueber die W{\"a}rmeleitungsf{\"a}higkeit
  von Fl{\"u}ssigkeiten}}.  \jt{Annalen der Physik und Chemie}  \bvol{25},
  \pg{337--357}, part 2; Part 1 appeared in vol.~18 (1883), pp.~79--94.

\bibitem[Harrandt {\em et~al.\/}(2024)Harrandt, Bazaikin, Huchet, Tihon \&
  Havlica]{HarrandtHavlica2024}
{\sc \au{Harrandt, V.}, \au{Bazaikin, Y.}, \au{Huchet, F.}, \au{Tihon, J.} \&
  \au{Havlica, J.}} \yr{2024}  \at{Theory for electrodiffusional wall shear
  stress measurement by directionally sensitive twin semicircular probe}.
  \jt{International Journal of Heat and Mass Transfer}  \bvol{235},
  \pg{126191}.

\bibitem[Harrandt {\em et~al.\/}(2025)Harrandt, Bazaikin, Huchet, Tihon \&
  Havlica]{HarrandtHavlica2025}
{\sc \au{Harrandt, V.}, \au{Bazaikin, Y.}, \au{Huchet, F.}, \au{Tihon, J.} \&
  \au{Havlica, J.}} \yr{2025}  \at{Generalized theory for wall shear stress
  measurement using circular-segment electrodiffusion probes}.
  \jt{International Communications in Heat and Mass Transfer}  \bvol{167}~(Part
  B),  \pg{109322}.

\bibitem[Harrandt {\em et~al.\/}(2026)Harrandt, Bazaikin, Huchet, Tihon \&
  Havlica]{HarrandtHavlica2026}
{\sc \au{Harrandt, V.}, \au{Bazaikin, Y.}, \au{Huchet, F.}, \au{Tihon, J.} \&
  \au{Havlica, J.}} \yr{2026}  \at{Analytical solution for the laminar wake
  behind a finite heat/mass sink}.  \jt{Journal of Fluid Mechanics}
  \bvol{1027},  \pg{A20}.

\bibitem[Harrandt {\em et~al.\/}(2023)Harrandt, Kramolis, Huchet, Tihon \&
  Havlica]{HarrandtHavlica2023}
{\sc \au{Harrandt, V.}, \au{Kramolis, D.}, \au{Huchet, F.}, \au{Tihon, J.} \&
  \au{Havlica, J.}} \yr{2023}  \at{The electrodiffusional theory for the wall
  shear stress measurement by two-strip probe}.  \jt{International Journal of
  Heat and Mass Transfer}  \bvol{212},  \pg{124287}.

\bibitem[Havlica {\em et~al.\/}(2021)Havlica, Kramolis \&
  Huchet]{HavlicaKramolis2021}
{\sc \au{Havlica, J.}, \au{Kramolis, D.} \& \au{Huchet, F.}} \yr{2021}  \at{A
  revisit of the electro-diffusional theory for the wall shear stress
  measurement}.  \jt{International Journal of Heat and Mass Transfer}
  \bvol{165},  \pg{120610}.

\bibitem[Huchet {\em et~al.\/}(2011)Huchet, Legentilhomme, Legrand, Montillet
  \& Comiti]{Huchet2011}
{\sc \au{Huchet, F.}, \au{Legentilhomme, P.}, \au{Legrand, J.}, \au{Montillet,
  A.} \& \au{Comiti, J.}} \yr{2011}  \at{Unsteady flows in milli- and
  microsystems: analysis of wall shear rate fluctuations}.  \jt{Experiments in
  Fluids}  \bvol{51}~(3),  \pg{597--610}.

\bibitem[L{\'e}v{\^e}que(1928)]{Leveque1928}
{\sc \au{L{\'e}v{\^e}que, M.~A.}} \yr{1928}  \at{Les lois de la transmission de
  chaleur par convection}.  \jt{Annales des Mines, M{\'e}moires}  \bvol{13},
  \pg{201--299, 305--362, 381--415}.

\bibitem[Liepmann(1958)]{Liepmann1958}
{\sc \au{Liepmann, H.~W.}} \yr{1958}  \at{A simple derivation of {L}ighthill's
  heat transfer formula}.  \jt{Journal of Fluid Mechanics}  \bvol{3}~(4),
  \pg{357--360}.

\bibitem[Lighthill(1950)]{Lighthill1950}
{\sc \au{Lighthill, M.~J.}} \yr{1950}  \at{Contributions to the theory of heat
  transfer through a laminar boundary layer}.  \jt{Proceedings of the Royal
  Society of London A}  \bvol{202}~(1070),  \pg{359--377}.

\bibitem[Mitchell \& Hanratty(1966)]{MitchellHanratty1966}
{\sc \au{Mitchell, J.~E.} \& \au{Hanratty, T.~J.}} \yr{1966}  \at{A study of
  turbulence at a wall using an electrochemical wall shear-stress meter}.
  \jt{Journal of Fluid Mechanics}  \bvol{26}~(1),  \pg{199--221}.

\bibitem[Newman(1967)]{Newman1967Report}
{\sc \au{Newman, John}} \yr{1967}  \bt{Extension of the {L}{\'e}v{\^e}que
  solution}. {\em Tech. Rep.\/} UCRL-17600.  \org{Lawrence Radiation
  Laboratory, University of California, Berkeley}.

\bibitem[Newman(1969)]{Newman1969}
{\sc \au{Newman, J.}} \yr{1969}  \at{Extension of the {L}{\'e}v{\^e}que
  solution}.  \jt{ASME Journal of Heat Transfer}  \bvol{91}~(1),
  \pg{177--178}.

\bibitem[Oldham \& Spanier(1974)]{OldhamSpanier1974}
{\sc \au{Oldham, K.~B.} \& \au{Spanier, J.}} \yr{1974} {\em The Fractional
  Calculus: Theory and Applications of Differentiation and Integration to
  Arbitrary Order\/},  \st{Mathematics in Science and Engineering},  \vol{vol.
  111}.  \publ{New York: Academic Press}, reprinted by Dover, 2006.

\bibitem[Podlubny(1999)]{Podlubny1999}
{\sc \au{Podlubny, I.}} \yr{1999} {\em Fractional Differential Equations\/},
  \st{Mathematics in Science and Engineering},  \vol{vol. 198}.  \publ{San
  Diego: Academic Press}.

\bibitem[Shah {\em et~al.\/}(2017)Shah, Lin \& Shaqfeh]{ShahLinShaqfeh2017}
{\sc \au{Shah, Preyas~N.}, \au{Lin, Tiras~Y.} \& \au{Shaqfeh, Eric S.~G.}}
  \yr{2017}  \at{Heat/mass transport in shear flow over a reactive surface with
  inert defects}.  \jt{Journal of Fluid Mechanics}  \bvol{811},  \pg{372--399}.

\bibitem[Shah \& London(1978)]{ShahLondon1978}
{\sc \au{Shah, R.~K.} \& \au{London, A.~L.}} \yr{1978} {\em Laminar Flow Forced
  Convection in Ducts\/}. {\em Advances in Heat Transfer\/} Supplement 1.
  \publ{New York: Academic Press}.

\bibitem[Sirkar \& Hanratty(1970)]{SirkarHanratty1970}
{\sc \au{Sirkar, K.~K.} \& \au{Hanratty, T.~J.}} \yr{1970}  \at{Relation of
  turbulent mass transfer to a wall at high {S}chmidt numbers to the velocity
  field}.  \jt{Journal of Fluid Mechanics}  \bvol{44}~(3),  \pg{589--603}.

\bibitem[Sobol{\'\i}k {\em et~al.\/}(1990)Sobol{\'\i}k, Wein, Gil \&
  Tribollet]{Sobolik1990}
{\sc \au{Sobol{\'\i}k, V.}, \au{Wein, O.}, \au{Gil, O.} \& \au{Tribollet, B.}}
  \yr{1990}  \at{Three-segment electrodiffusion probes for measuring velocity
  fields close to a wall}.  \jt{Experiments in Fluids}  \bvol{9}~(1--2),
  \pg{43--48}.

\bibitem[Son \& Hanratty(1969)]{SonHanratty1969}
{\sc \au{Son, J.~S.} \& \au{Hanratty, T.~J.}} \yr{1969}  \at{Velocity gradients
  at the wall for flow around a cylinder at {R}eynolds numbers from $5\times
  10^3$ to $10^5$}.  \jt{Journal of Fluid Mechanics}  \bvol{35}~(2),
  \pg{353--368}.

\bibitem[Spence \& Brown(1968)]{SpenceBrown1968}
{\sc \au{Spence, D.~A.} \& \au{Brown, G.~L.}} \yr{1968}  \at{Heat transfer to a
  quadratic shear profile}.  \jt{Journal of Fluid Mechanics}  \bvol{33}~(4),
  \pg{753--773}.

\bibitem[Wein \& Wichterle(1989)]{WeinWichterle1989}
{\sc \au{Wein, O.} \& \au{Wichterle, K.}} \yr{1989}  \at{Theory of segmented
  electrodiffusion probes: the effect of insulating insertions}.
  \jt{Collection of Czechoslovak Chemical Communications}  \bvol{54},
  \pg{3198--3212}.

\bibitem[Wors{\o}e-Schmidt(1967)]{WorsoeSch1967}
{\sc \au{Wors{\o}e-Schmidt, P.~M.}} \yr{1967}  \at{Heat transfer in the thermal
  entrance region of circular tubes and annular passages with fully developed
  laminar flow}.  \jt{International Journal of Heat and Mass Transfer}
  \bvol{10}~(4),  \pg{541--551}.

\end{thebibliography}
